\documentclass[pra,twocolumn,showpacs,aps,floatfix,amsmath,amssymb,amsfonts]{revtex4}
\pdfoutput=1
\usepackage{amsmath}
\usepackage{graphicx}
\usepackage{color}

\newcommand{\mbf}{\mathbf}

\begin{document}
\title{Light scattering from ultracold gases in disordered optical lattices}
\author{Krzysztof Jachymski and Zbigniew Idziaszek}
\affiliation{Faculty of Physics, University of Warsaw, Ho{\.z}a 69,
00-681 Warsaw, Poland}

\date{\today}

\begin{abstract}
We consider a gas of bosons in a bichromatic optical lattice at finite temperatures. As the amplitude of the secondary lattice grows, the single-particles eigenstates become localized. We calculate the canonical partition function using exact methods for the noninteracting and strongly interacting limit and analyze the statistical properties of the superfluid phase, localized phase and the strongly interacting gas. We show that those phases may be distinguished in experiment using off-resonant light scattering.
\end{abstract}

\maketitle

\section{Introduction}

In 1958 Anderson proposed a mechanism that explained the absence of conductance in certain types of media \cite{Anderson1958}. By considering the propagation of matter waves through a medium containing randomly distributed impurities, he discovered that under some conditions the single-particle states become exponentially localized and the propagation is blocked. Anderson argued that the effect occurs due to destructive interference of waves, therefore it can be observed both for quantum matter waves and for classical waves. So far Anderson localization (AL) has been reported for acoustic waves \cite{Tiggelen}, microwaves \cite{Chabanov}, and light \cite{Wiersma}. It has been also realized in ultracold atomic gases confined in optical random potentials \cite{Inguscio,Aspect}. In comparison to other physical systems, ultracold atoms offer unique possibilities of control of both the shape of the external potential and atomic interactions \cite{Bloch2008}. Disordered potentials can be generated either by use of the laser-speckle pattern \cite{AspectOld,AspectJourn} or a two-color optical lattice \cite{InguscioPRA,Guarrera2007}. In the former case the potential is truly random, while in the latter the system is quasi-periodic.

Localization properties depend strongly on the form of the potential and the dimensionality of the system. It is known that in one dimension AL is present even for arbitrarily small random potential \cite{Thouless,Gang4,Proof}. However, for a pseudo-random potential, such as the two-color optical lattice, there is a transition to localized states only at certain disorder strength \cite{Aubry1980}. 

One of the most important challenges is to understand the interplay between the disorder and interactions. Adding the interactions may lead to novel quantum phases, such as the Bose glass phase, which appears in addition to superfluid and Mott insulator phases \cite{Fisher1989,Lewenstein2003}. The phase diagram of such systems can be complicated and hard to obtain theoretically (see e.g. \cite{Roux2008,Fontanesi2008}). Another challenge is to include the role of finite temperature, which excites the particles and suppresses localization.

Detecting and studying the correlation properties of the quantum phases in experiment is not easy as well. The usual detection scheme is to switch the trap off and let the atoms expand for a certain time, measuring the interference pattern \cite{BlochNature}. However, the information provided by this method is very limited \cite{Burnett2003}, in particular in case of disordered lattices. Therefore new experimental schemes has been developed, e.g. based on the noise correlation measurements \cite{BlochNoise,Demler}, or observing the far-field pattern in coherent light-scattering \cite{Bloch2011}. So far methods based on atom-light interactions have been proposed for several various situations, such as detection of the Bose condensed phase \cite{Lewenstein1993}, condensate fluctuations \cite{Idziaszek}, superfluidity in Fermi gases \cite{Zhang1999} and for detection of quantum phases in optical lattices \cite{Mekhov2007,Lakomy,Menotti2010,Douglas2011}. The spectrum of weak and far-detuned light carries information about the static structure factor of the investigated system \cite{Burnett2003,Lakomy,Douglas2011}, which contains information about the density-density correlations and the energy spectrum. The structure factor is also accessible via Bragg diffraction \cite{Burnett2003,Bragg}. The phases of strongly correlated systems can be also detected using quantum-noise-limited polarization spectroscopy \cite{LewensteinNature}.

In this work we investigate the finite temperature properties of a one-dimensional Bose gas in a bichromatic optical lattice. We focus on two limits when the partition function can be found exactly. One is the noninteracting gas described by the Aubry-Andre Hamiltonian and the other is strongly interacting gas with negligible inter-well tunneling. We calculate the mean, fluctuations and correlations of occupation numbers in the lattice wells. In addition, for the ideal gas we examine the condensate fraction and its fluctuations. In the second part of the paper we analyze the angular distribution of light scattered from the ideal and strongly interacting gas. We show that similarly to regular optical lattice the light can be used to discriminate different phases existing in disordered systems.

Our paper is organized as follows. In section II we introduce the Aubry-Andre model, review its single-particle properties and analyze the energy spectrum. Section III is devoted to the statistical properties of the gas in two-color optical lattice and the impact of temperature on the localization properties. We consider the strongly interacting gas in section IV. Section V studies the properties of the off-resonant light scattered on noninteracting and strongly interacting gases in disordered potentials systems and discusses distinguishability of different phases with this method. Section VI presents the conclusions and three appendices give some technical details on exact calculations of the statistical quantities in the canonical ensemble.

\section{Bose-Hubbard model for disordered system}

We consider a 1D Bose gas in the optical lattice in the presence of a weak disorder. We assume that the gas is sufficiently cold and its dynamics takes place only in the lowest Bloch band. In such a case the dynamics is governed by the Bose-Hubbard Hamiltonian~\cite{Jaksch1998}
\begin{equation}
H = - J \sum_{\langle i,j \rangle} g_i^\dagger g_j + \frac{U}{2} \sum_i n_i (n_i - 1) +
\sum_i n_i \varepsilon_i
\label{hamiltonian}
\end{equation}
with an additional term describing the on-site energies $\varepsilon_i$ due to the disorder \cite{Lewenstein2003,Roux2008}. Here, $g_i$ and $g_i^\dagger$ are the annihilation and creation operators of bosons at lattice site $i$, respectively, $n_i = g_i^\dagger g_i$ is the particle number operator at site $i$ and $J$ and $U$ are the energy scales corresponding to the tunneling between wells and the on-site interaction. In our approach we consider disorder induced by a
bichromatic optical lattice potential \cite{Aubry1980}
\begin{equation}
V(x) = s_1 E_{r1} \sin^2 (k_1 x) + s_2 E_{r2} \sin^2 (k_2 x + \phi).
\label{potencjal}
\end{equation}
Here, $k_i = 2\pi / \lambda_i$ are the wave numbers of the light beams creating the standing wave, $E_i = h^2 / 2 m \lambda_i ^2$ are the recoil energies, $s_i$ are the heights of the two lattice potentials in units of recoil energies, $\phi$ is a relative phase between two laser beams and $m$ is the atom mass. We will denote the ratio of the wave numbers $k_1/k_2$ as $\beta$. For $s_1 E_{r1} \gg s_2 E_{r2}$ the first lattice generates the periodic structure of the potential, while the second lattice generates weak quasi-periodic modulation of the potential wells. In this case
\begin{equation}
\varepsilon_i = \Delta \cos(2 \pi \beta i + 2 \phi).
\end{equation}
Here, $\Delta$ is the measure of the disorder strength. It can be expressed in terms of the Wannier states $w(x)$ localized in the wells of the first lattice~\cite{Modugno2009}: $\Delta = E_{r1} \frac{s_2 \beta^2}{2} \int{d\xi} \cos(2\beta \xi) \left|w(\xi)\right|^2$.

The case of an ideal gas $U=0$ at zero temperature has been extensively studied in the literature \cite{Aubry1980,Modugno2009,Aulbach}. The disorder introduced in this Hamiltonian is pseudorandom, and it has been shown that even in one dimension the eigenstates of the single-particle Hamiltonian are not localized for low $\Delta$, in contrast to the standard Anderson localization~\cite{Proof}. Instead, if $\beta$ is an irrational Diophantine number, there is a transition from extended to localized states. In particular case of $\beta = (\sqrt{5}-1)/2$ the transition occurs at $\Delta=2\,J$, which is a self dual point~\cite{Aubry1980,Jitomirskaya1999}.
In practice, the system of atoms in an optical lattice has a finite size, therefore it is sufficient that $\beta$ is a rational number and the periodicity of the on-site energy modulation $\varepsilon_n$ is larger than the system size.

\section{Ideal gas}

In this section we consider the statistical properties of an ideal gas confined in a quasi-periodic potential. We start by studying the single particle properties, then we study the statistics of the gas at finite temperatures.

\subsection{Single-particle states}

In case of an ideal gas ($U=0$) the elementary excitations of the Hamiltonian~\eqref{hamiltonian}
\begin{equation}
H = - J \sum_{\langle k,l \rangle} g_k^\dagger g_l + \Delta \sum_k n_k \cos(2 \pi \beta k)
\label{hamiltonian1}
\end{equation}
have been already studied by Aubry and Andre \cite{Aubry1980}. As the Hamiltonian is quadratic it can be easily diagonalized in the basis of states describing atoms localized in a single potential well. In this way the single-particle states $|\Psi_n\rangle$ and corresponding single-particle energies $\epsilon_n$ can be expressed as
\begin{align}
H |\Psi_n\rangle & = \epsilon_n |\Psi_n\rangle, \\
|\Psi_n\rangle & = \sum_i c^n_i g^\dag_i |\Omega\rangle,
\end{align}
where $|\Omega\rangle$ denote the vacuum state, and $c^n_i$ are expansion coefficients. We calculate the energy spectrum and eigenstates of the model numerically. In original formulation of Aubry-Andre model $\beta$ is irrational and the system is quasiperiodic. In this case it is impossible to use periodic boundary conditions. In practice, however, it is sufficient to use a rational $\beta$ with sufficiently large system so that the periodicity of the energy modulation $\varepsilon_m$ is equal to the system size. A practical way to do that is to use Fibonacci numbers $f_i$, setting $\beta=f_{n-1}/f_n$ and $M$ (the number of lattice sites) to $f_n$. Most of the numerical results in this work are obtained using $M=144$ and $\beta=89/144$, which is a sufficiently good approximation.

\begin{figure}
\centering
\includegraphics[width=0.48\linewidth]{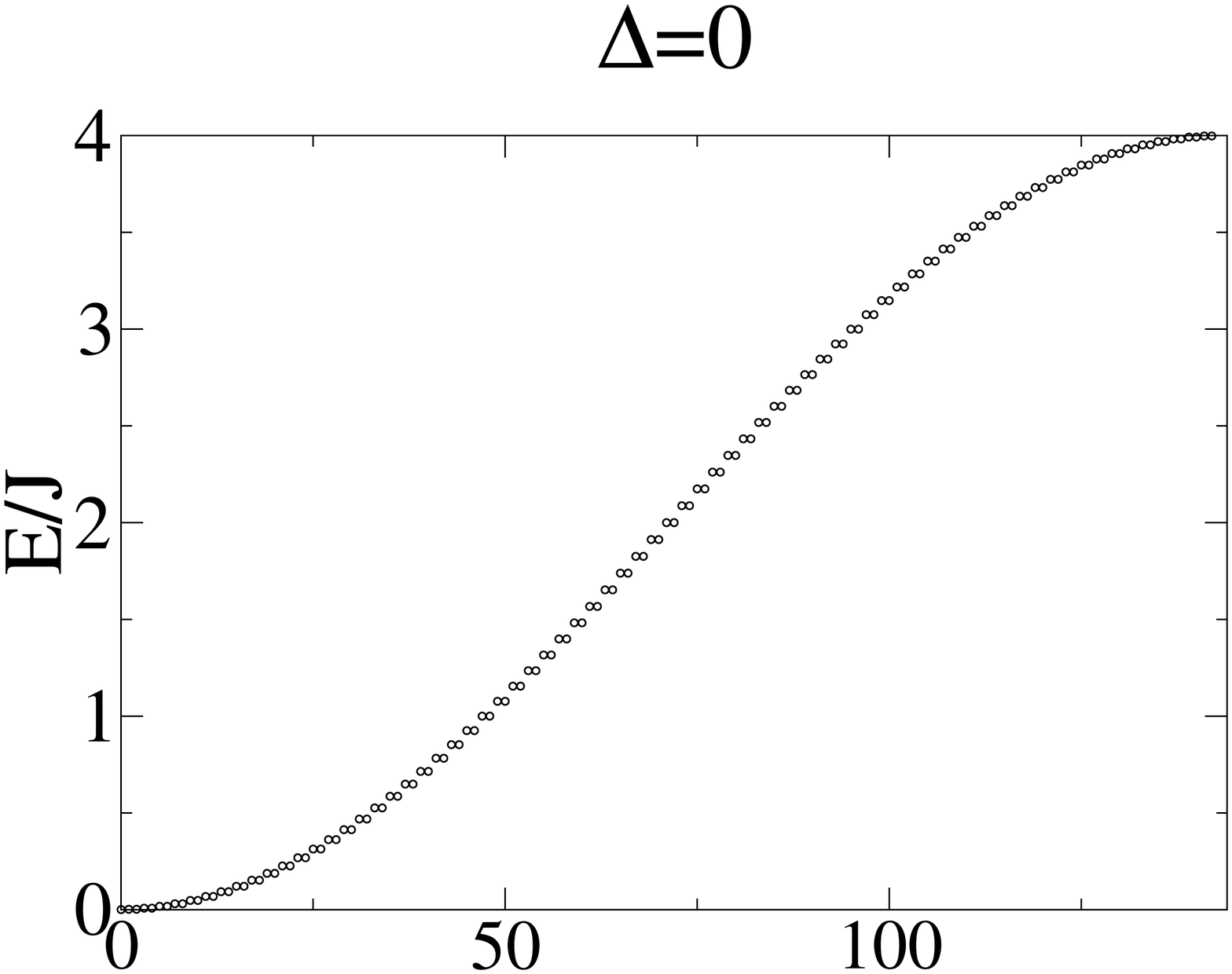}
\includegraphics[width=0.48\linewidth]{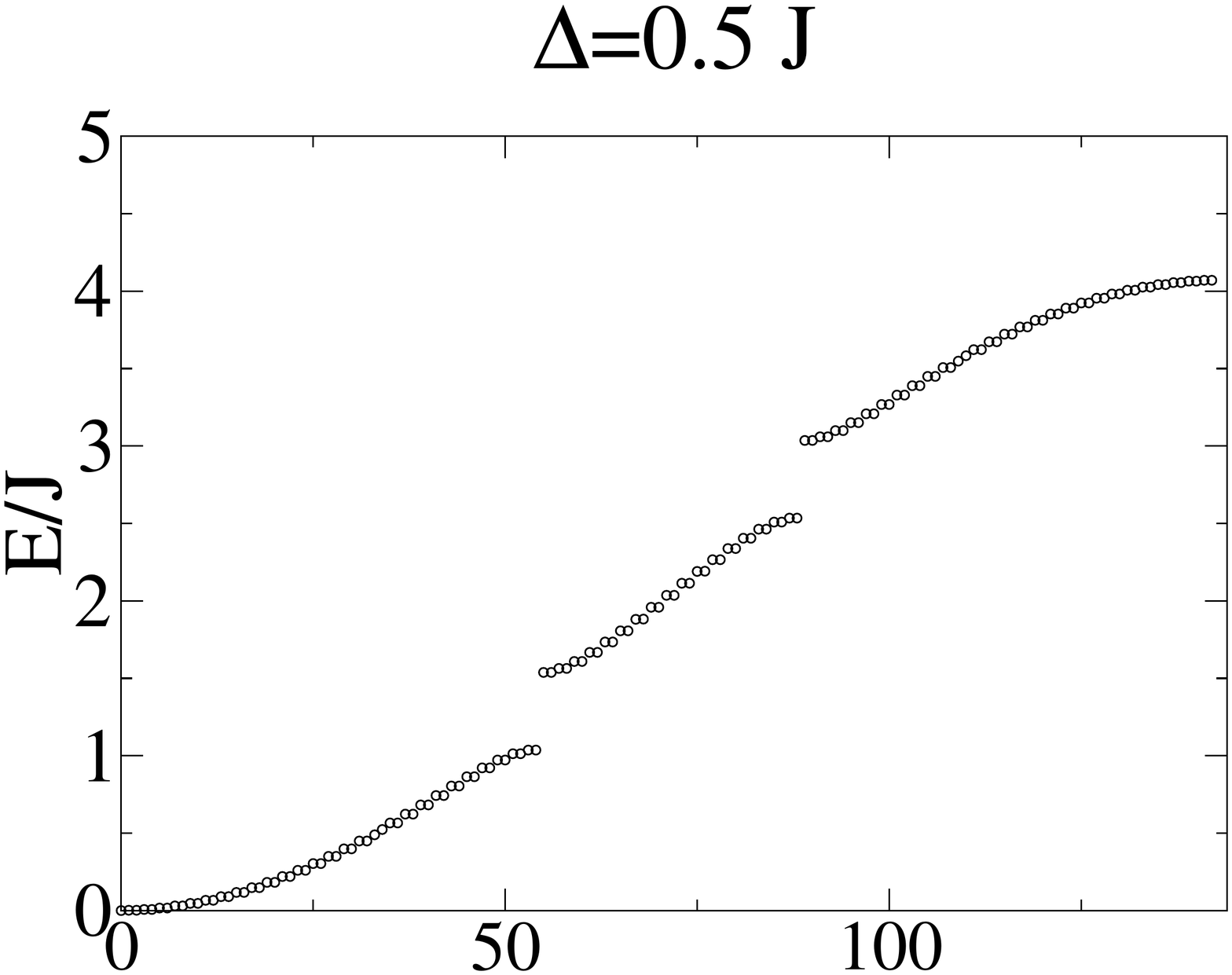}\\
\includegraphics[width=0.48\linewidth]{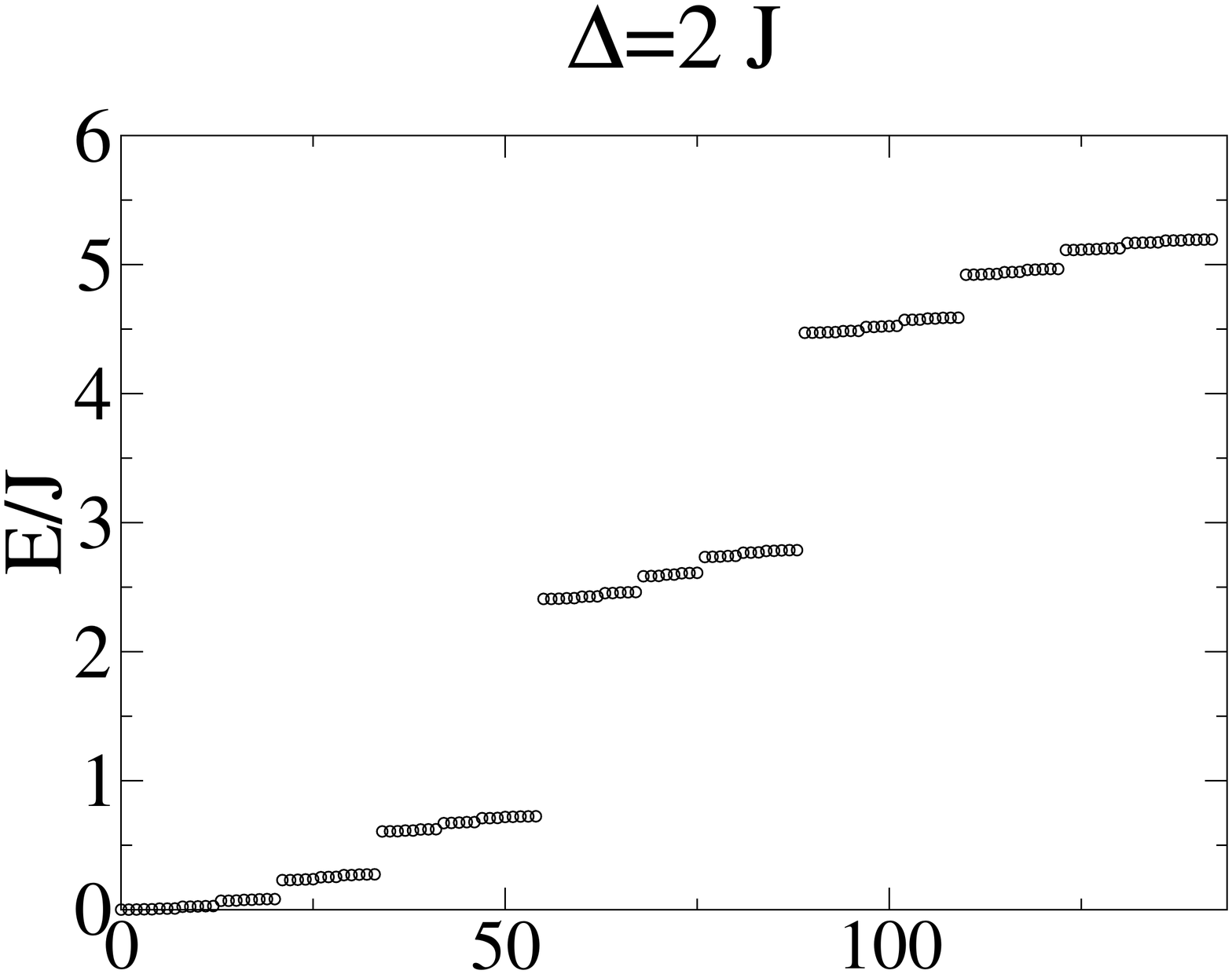}
\includegraphics[width=0.48\linewidth]{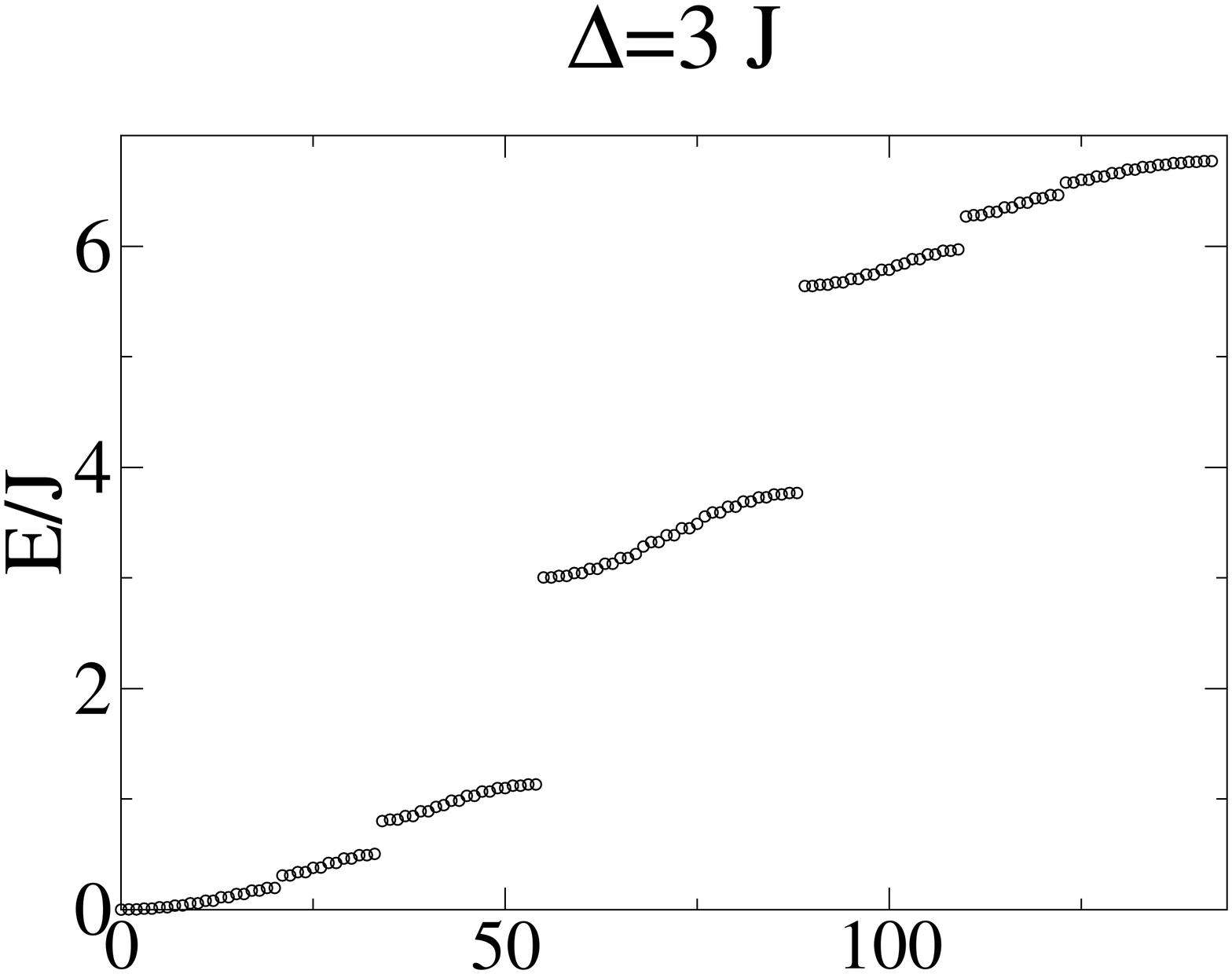}
\caption{Energy spectrum of the Aubry-Andre Hamiltonian \eqref{hamiltonian1} for $M = 610$ lattice sites and for different values of the disorder amplitude $\Delta$.}
\label{pic_spec}
\end{figure}

\begin{figure}
       \centering
       \includegraphics[width=0.48\linewidth]{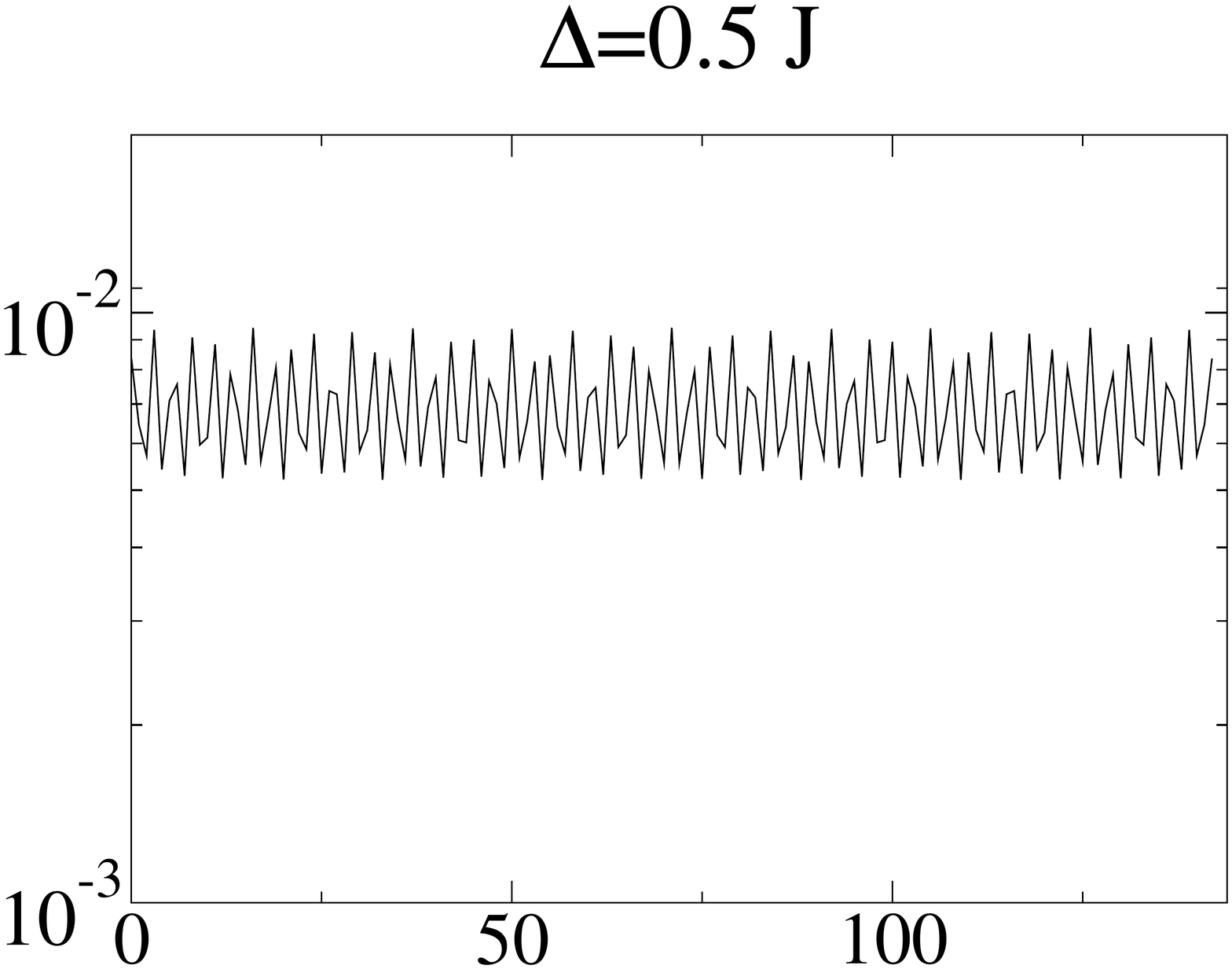}
       \includegraphics[width=0.48\linewidth]{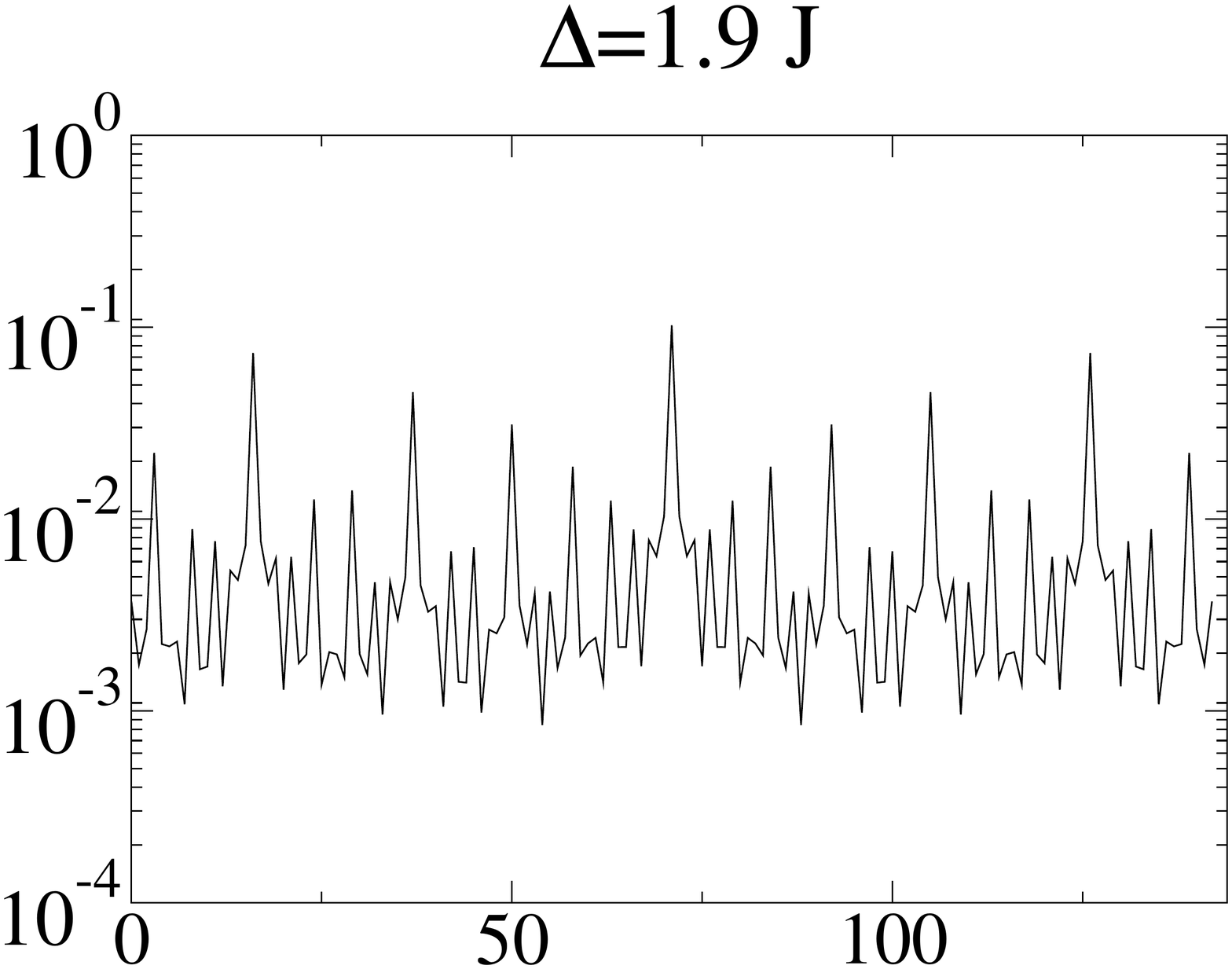}\\
       \includegraphics[width=0.48\linewidth]{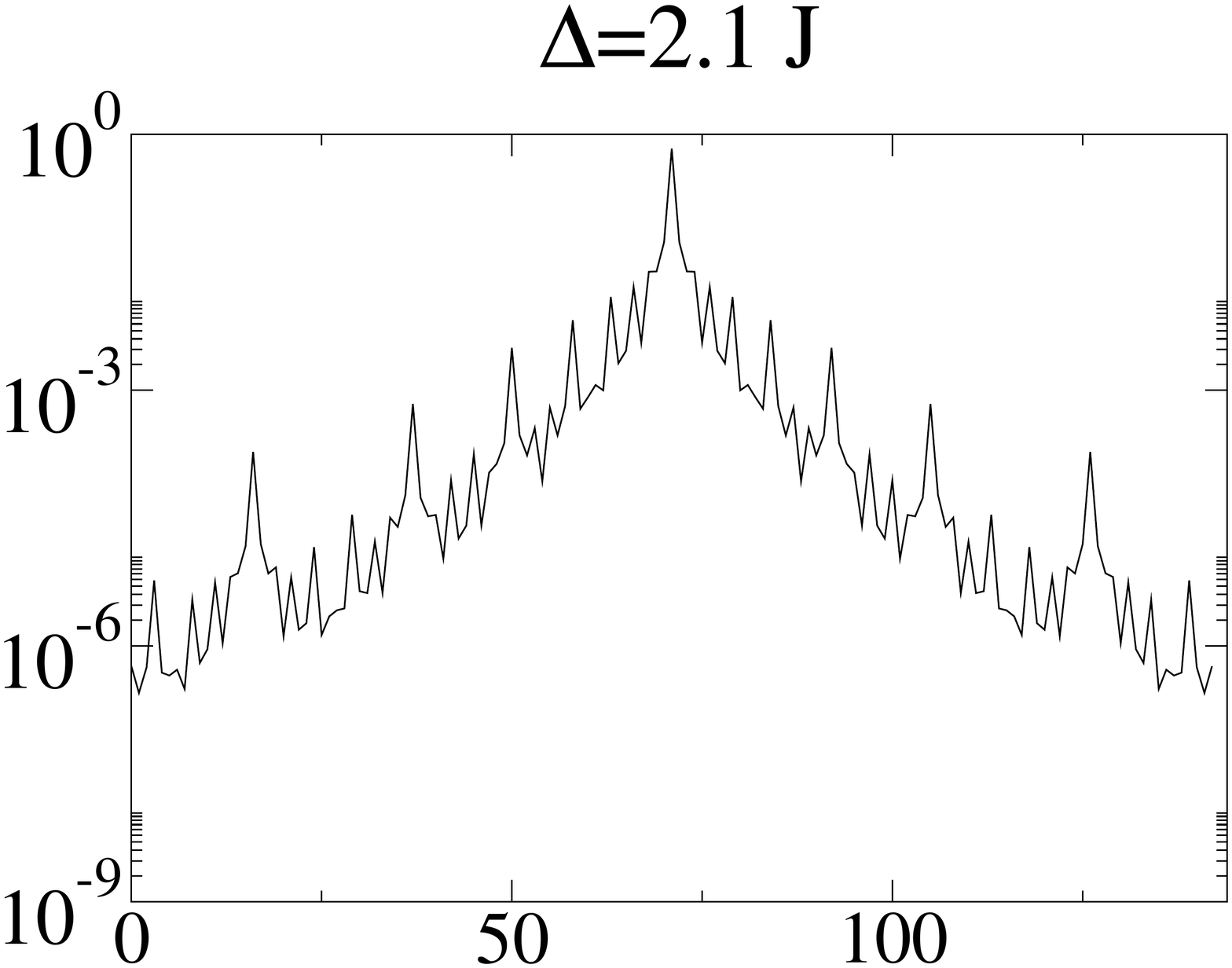}
       \includegraphics[width=0.48\linewidth]{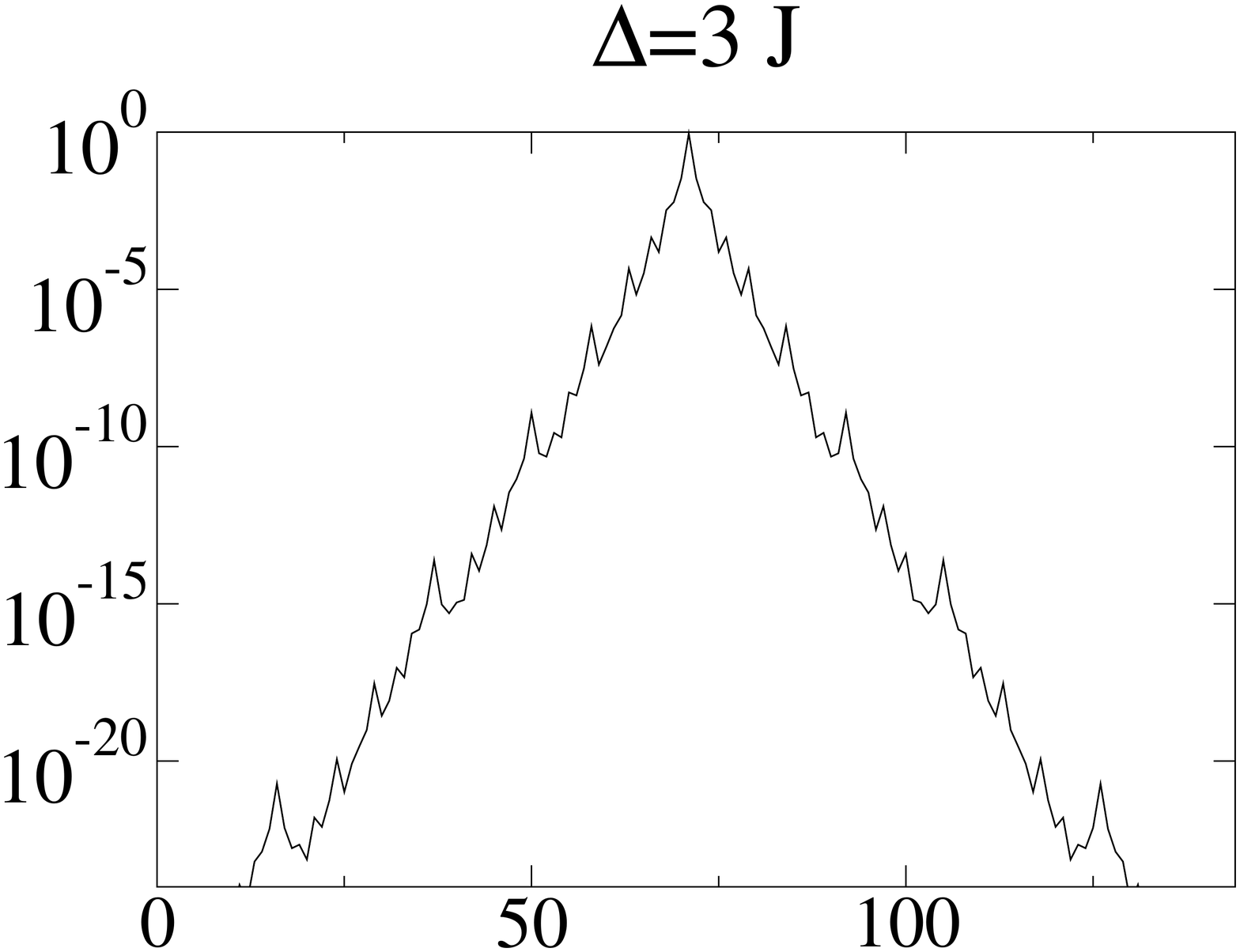}
       \caption{Probability distribution $|c_i^0|^2$ of finding atoms in different lattice sites for atoms in the ground state and for various disorder strengths $\Delta$. We use logarithmic scale for $y$-axis to show exponential localization.}
       \label{groundstate}
\end{figure}

\begin{figure}
\centering
\includegraphics[width=0.48\linewidth]{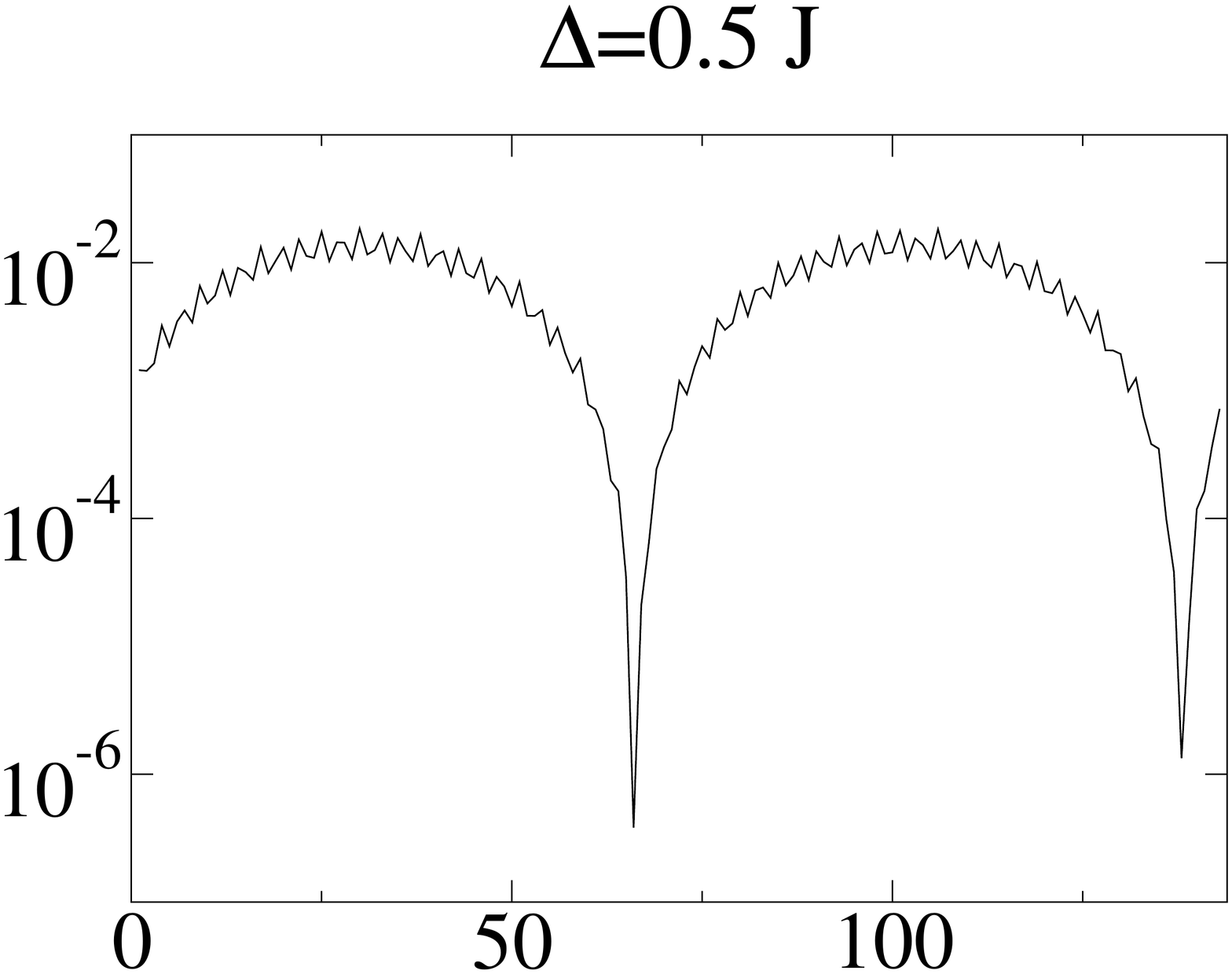}
\includegraphics[width=0.48\linewidth]{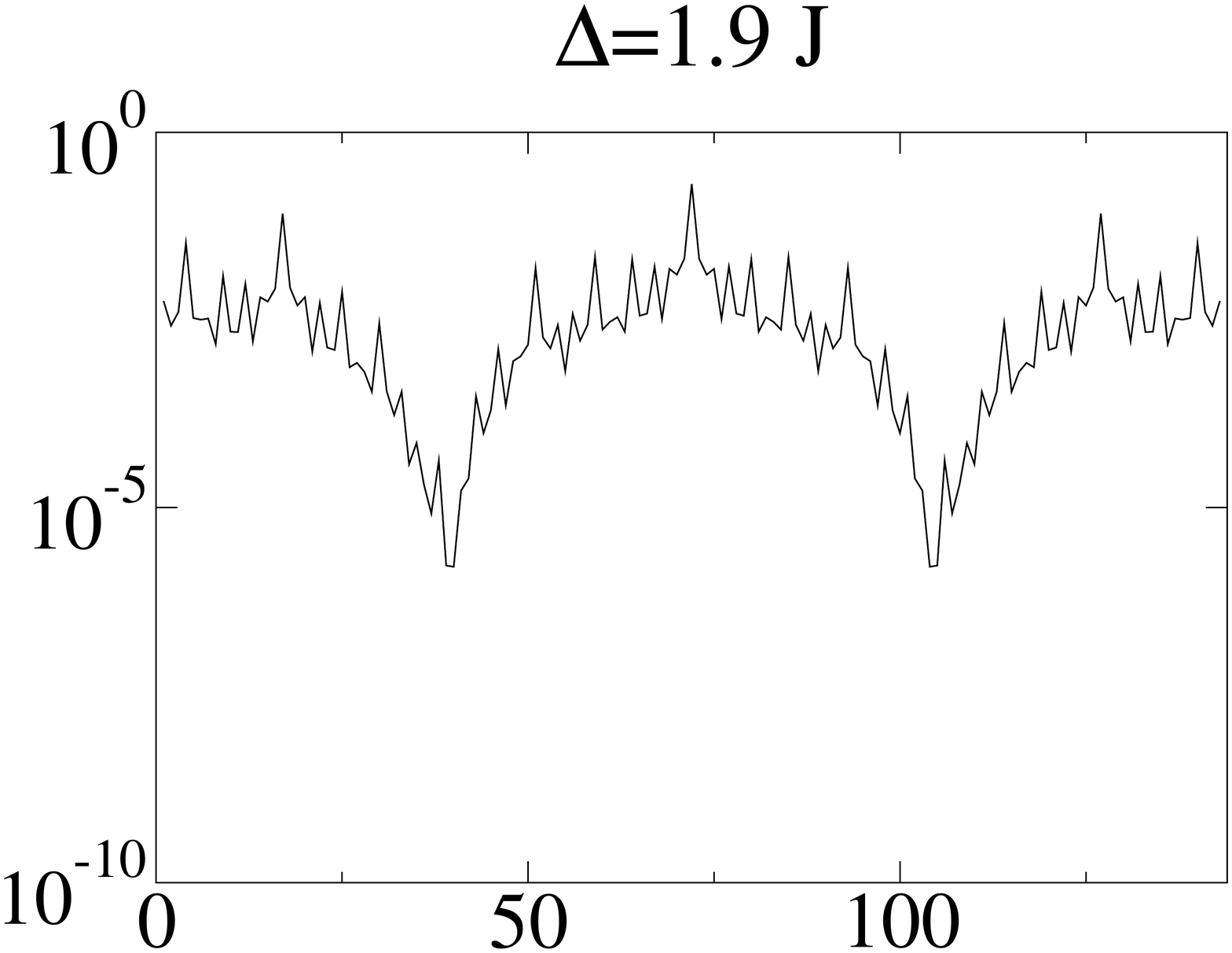}\\
\includegraphics[width=0.48\linewidth]{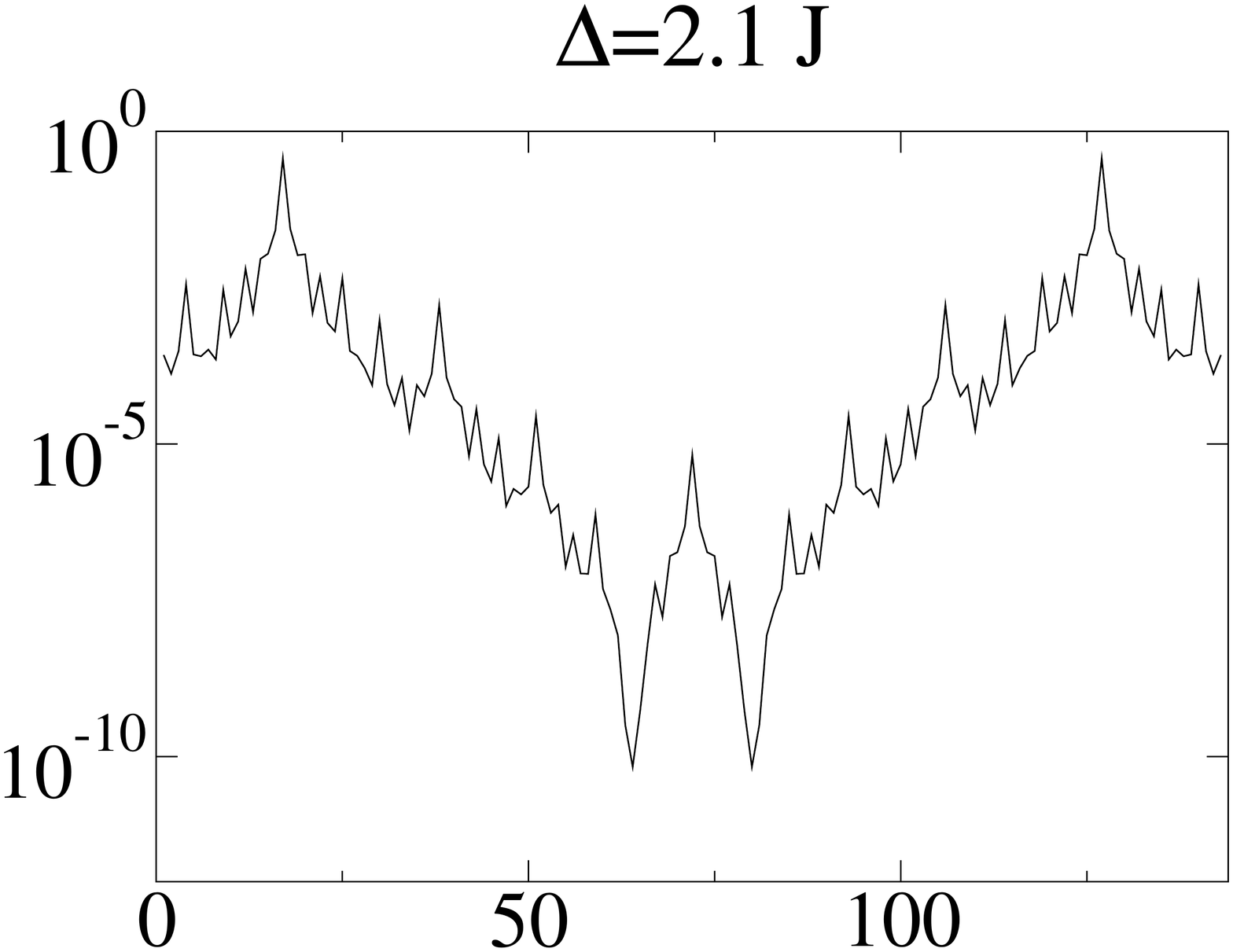}
\includegraphics[width=0.48\linewidth]{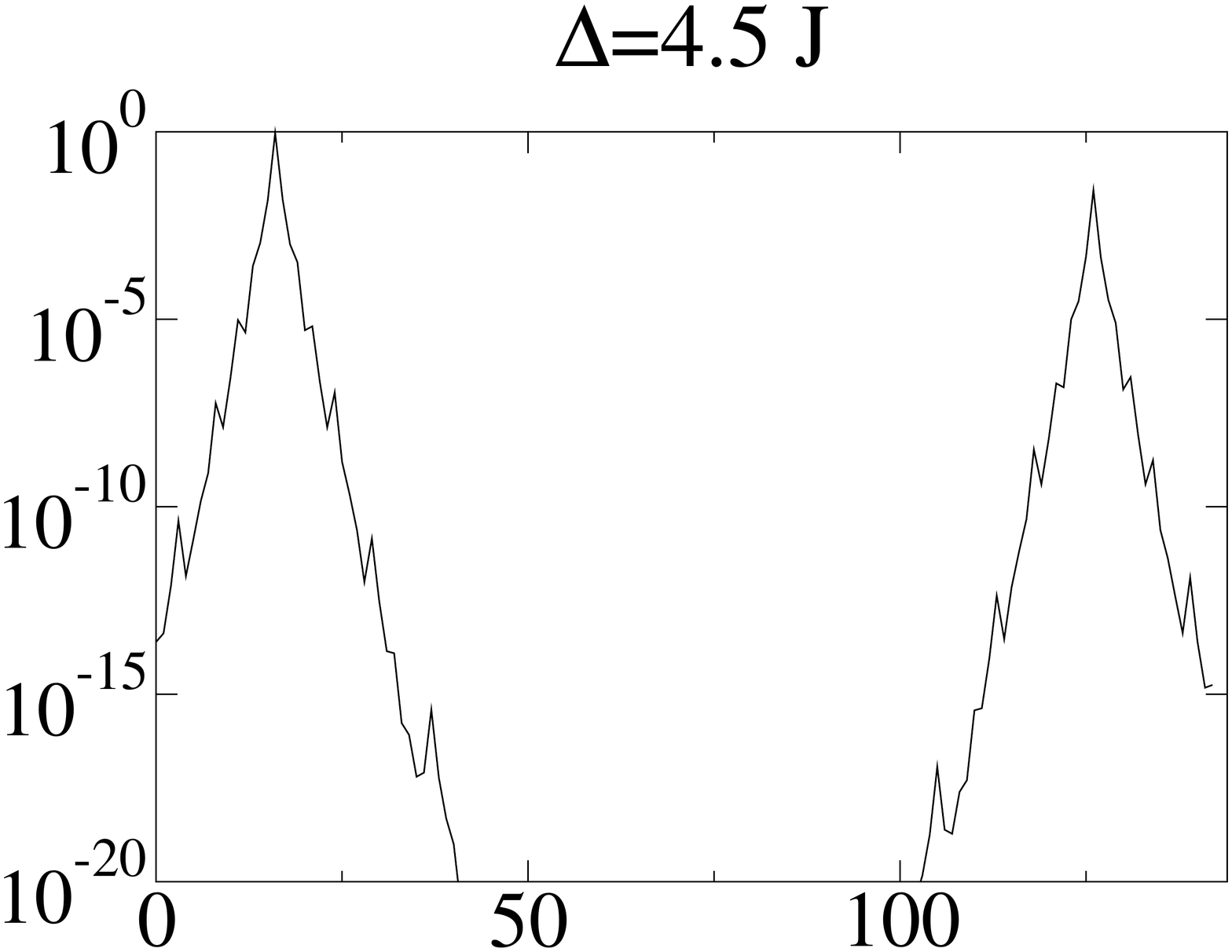}
\caption{Similar to Fig.~\ref{groundstate}, but for the first excited state. For $\Delta=4.5\,J$ the dominating left peak is two orders of magnitude stronger than the other.}
\label{pic_ex1}
\end{figure}

Figure~\ref{pic_spec} shows the energy spectrum for different strengths of the disorder. When $\Delta = 0$ the energy spectrum is just a lowest Bloch band, like in the Bose-Hubbard model. When the disorder increases, some energy levels tend to form groups, separated from the others by energy gaps. The effect is strongest for $\Delta = 2\,J$, for even larger $\Delta$ the spectrum becomes more regular again. The behavior of the ground and the first excited state for different values of $\Delta$ is shown on Figs.~\ref{groundstate} and \ref{pic_ex1}. The ground state for $\Delta=0$ is almost uniformly distributed over the whole lattice. When $\Delta$ gets larger the probability distribution becomes nonuniform and some lattice sites are favored. Finally for $\Delta=2$ the ground state become exponentially localized in a single site. In contrast, for the same value of $\Delta$ the first excited state exhibits two maxima localized in two distant lattice wells. For even larger $\Delta$, however, the first excited state becomes localized in a single lattice well. We have verified that a similar behavior occurs for higher excited states.

\subsection{Statistical properties at finite temperatures}

We now examine the properties of noninteracting gas in a bichromatic lattice at finite temperature. As the grand-canonical ensemble predicts unphysically large condensate fluctuation at ultralow temperatures when the ground state is macroscopically populated, the ultracold ideal gas of atoms has to be described either in the microcanonical or the canonical ensemble \cite{Politzer1996,Gajda1997,Navez}. The former one assumes the perfect isolation of the system from the environment, while the latter one assumes that the system is in contact with a heat bath of certain temperature: $k_B T=1/\beta$. Both ensembles correctly describe the fluctuations and correlations of an ideal gas at low temperatures. In our approach we apply the canonical ensemble. Its partition function $Z(\beta,N)$ can be defined as
\begin{equation}
Z(\beta,N)=\sum_{n_1=0}^\infty{}\sum_{n_2=0}^\infty{}\ldots\sum_{n_\infty=0}^\infty{} e^{-\beta \sum_\nu{n_\nu \epsilon_\nu}}\delta_{(\sum_i{n_i},N)},
\end{equation}
where $n_i$ denote the number of particles occupying the eigenstate with energy $\epsilon_i$, and $N$ is the total number of particles. The presence of a discrete delta function $\delta$ assures that only partitions with the total number of particles equal to $N$ contribute to the sum.
For a noninteracting system the partition function and all the other statistical quantities may be computed using the recurrence formulas (see Appendix A for details), based on the formula obtained in \cite{Weiss97}:
\begin{equation}
Z(\beta,N)=\sum_{n=1}^N{}\sum_\nu{e^{-n\beta\epsilon_\nu}Z(\beta,N-n)},
\label{recurrence}
\end{equation}
where we should take $Z(\beta,0)=1$.

\subsection{Ground state population behaviour}

\begin{figure}
\centering
\includegraphics[width=0.47\textwidth]{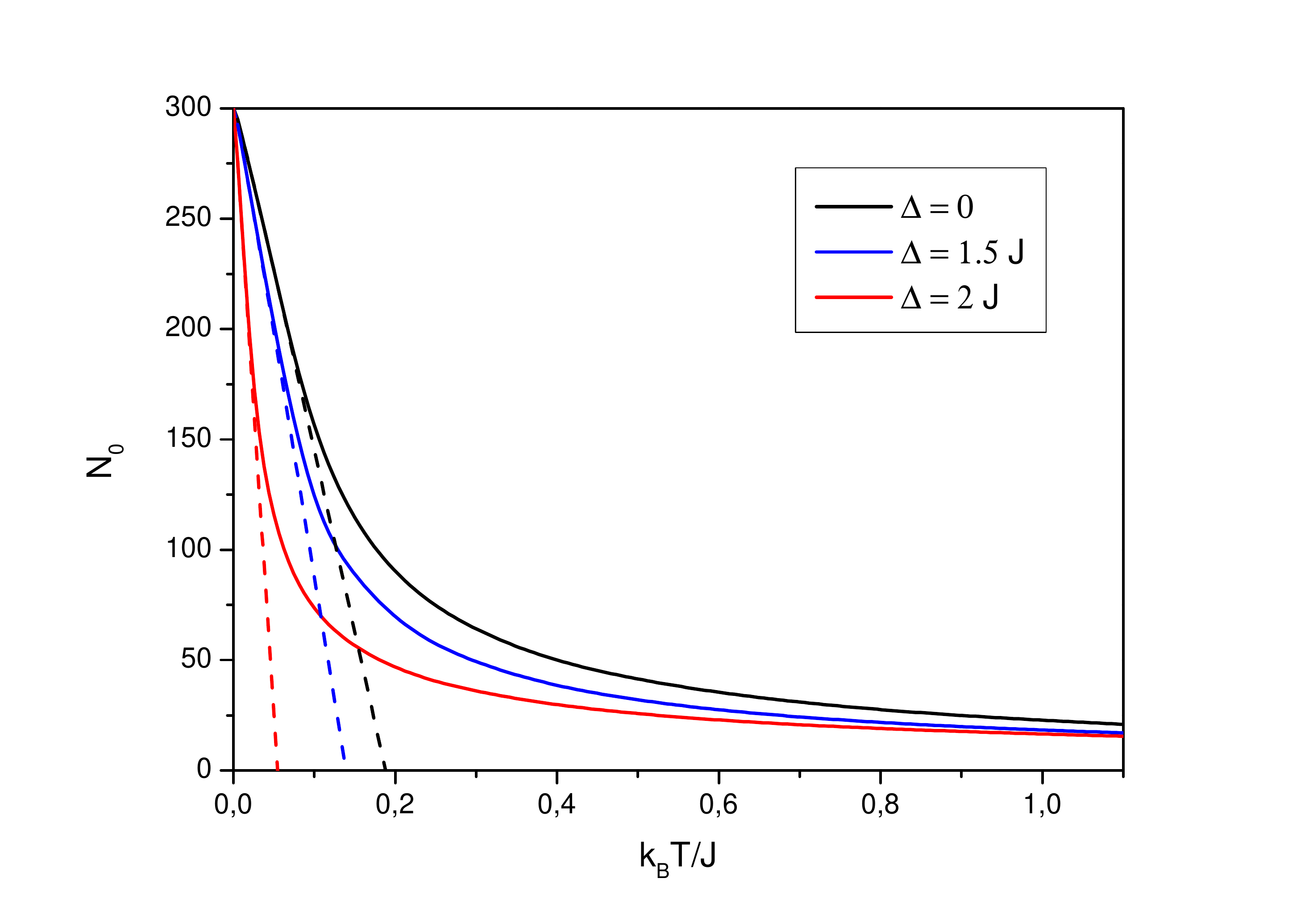}
\includegraphics[width=0.47\textwidth]{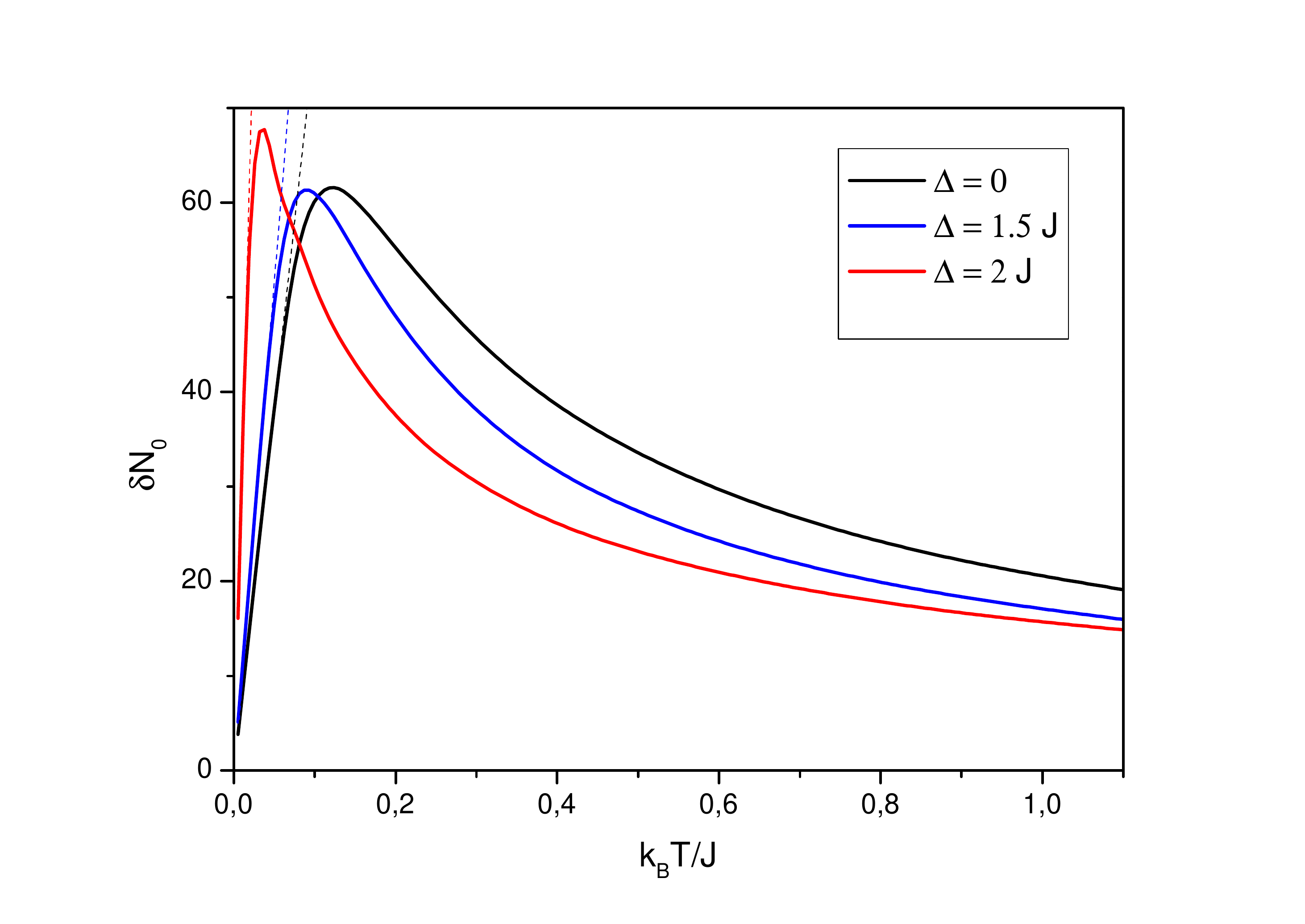}
\caption{Top: number of ground state atoms obtained in canonical ensemble for different values of $\Delta$, compared with the analytical model predictions. Bottom: fluctuations of the number of ground state atoms. $N=300$.}
\label{canvsMD}
\end{figure}

Having calculated the partition function, we can get the ground state population and its fluctuations. The numerical results are presented on Figure \ref{canvsMD}. First we observe the growth of fluctuations. However, as the ground state population decreases with temperature, so should its fluctuations. The maximum of $\delta N_0$ occurs at certain characteristic temperature which depends on disorder strength. We have developed an analytical model to explain this behaviour and to give some estimate on the characteristic temperature of the maximum of fluctuations. The shape of the energy spectrum for the lowest states may be approximated by a parabola. For an ideal gas with parabolic energy spectrum $\epsilon_n = a n^2$ all the statistical quantities can be calculated analytically. In one dimension one can introduce some characteristic temperature $T_c=6aN/\pi^2$, which determines the regime when the ground state becomes macroscopically populated (see Appendix B for details). Below $T_c$ the ground-state occupation number can be calculated using, for instance, the technique of the Maxwell-Demon ensemble \cite{Navez,Grossmann97}.
This yields
\begin{equation}
\frac{\left\langle N_0 \right\rangle}{N}\approx1-\frac{T}{T_c}.
\end{equation}
Above the characteristic temperature too many atoms become excited and the Maxwell-Demon method is not applicable. We observe that the model is reliable below temperature at which about a half of the particles become excited. The model based on the Maxwell-Demon ensemble gives also the correct value of the characteristic temperature $T_c$ at which the fluctuations reach the maximum.

\subsection{Mean number and fluctuations}

\begin{figure}
\centering
\includegraphics[width=0.45\textwidth]{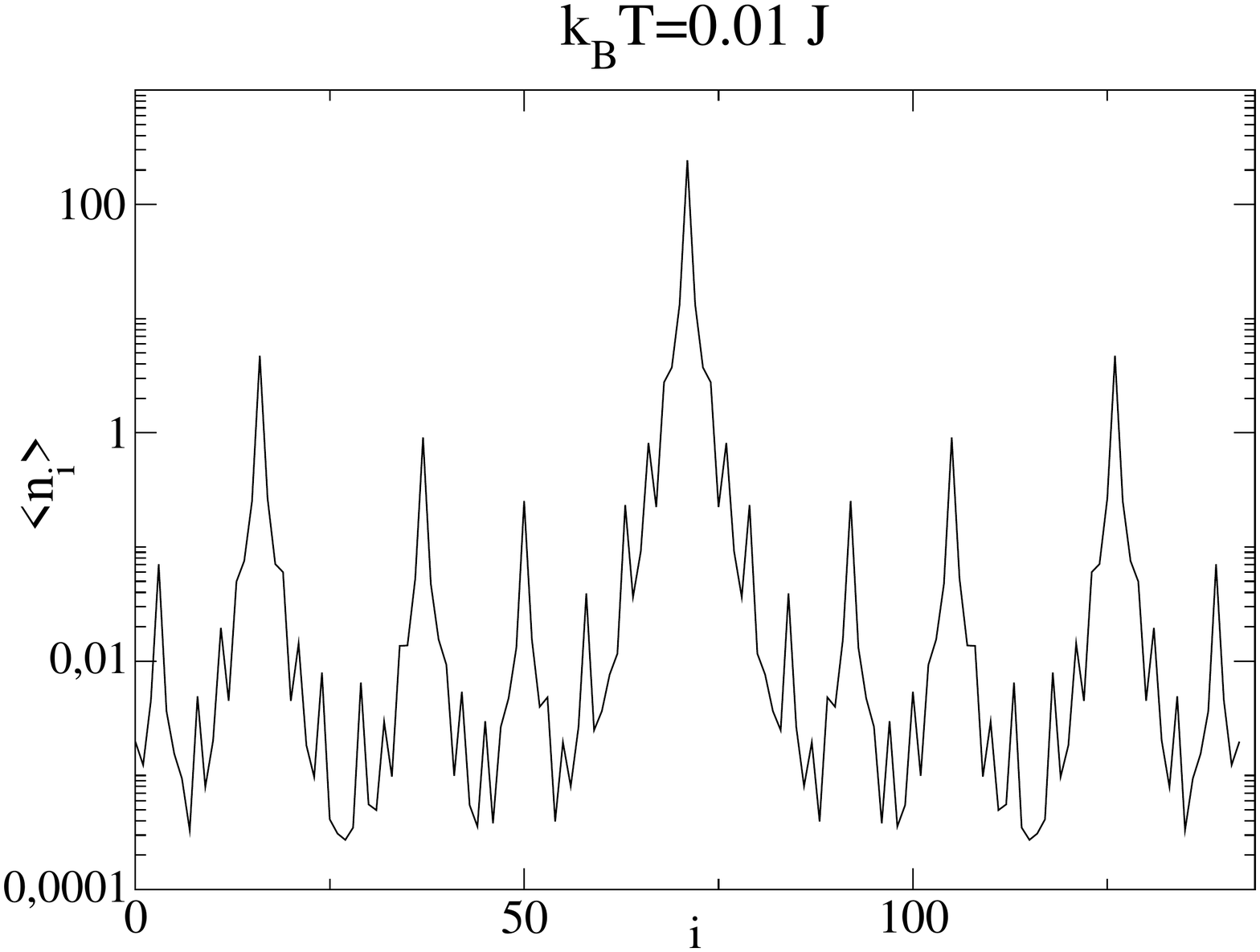}
\includegraphics[width=0.45\textwidth]{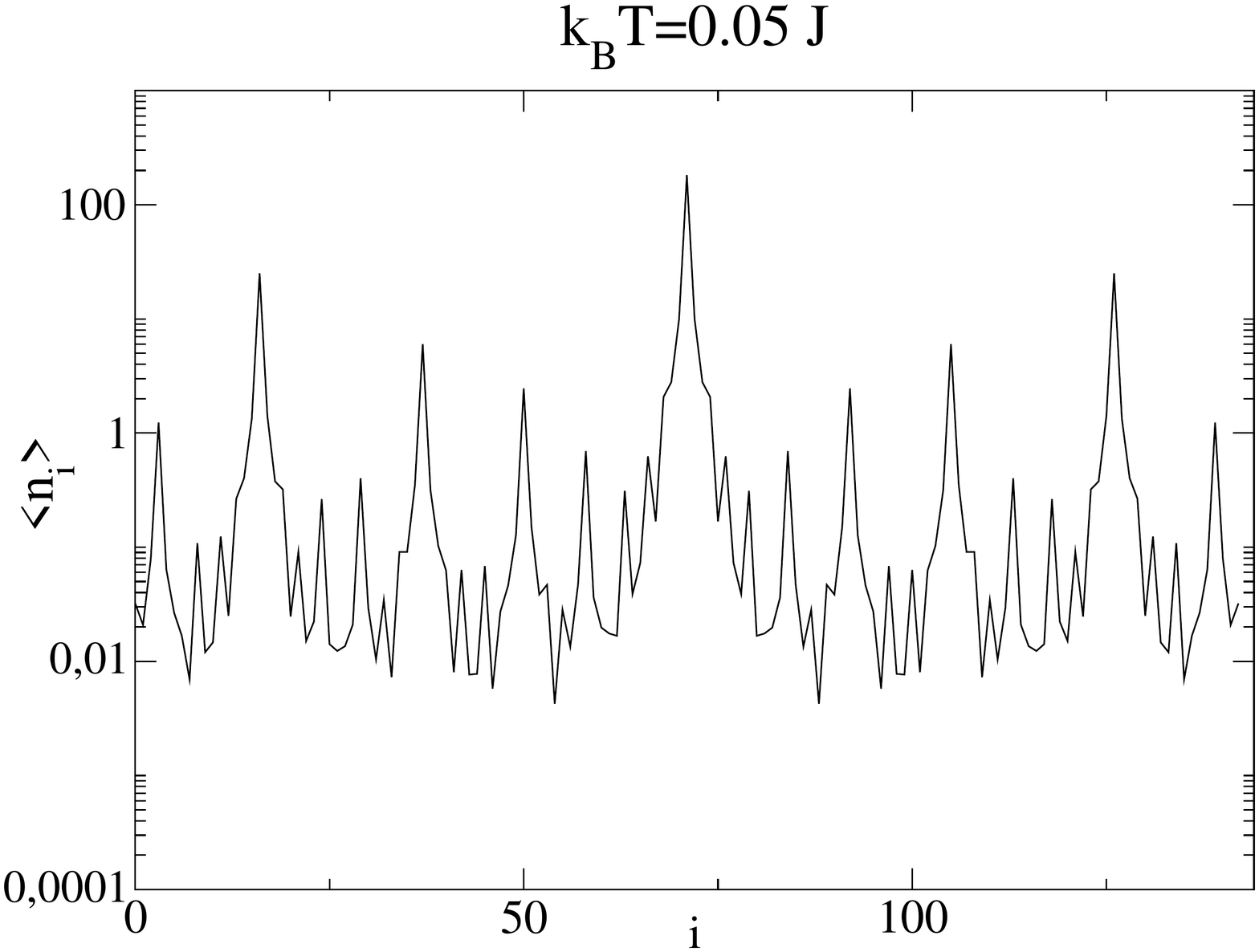}
\includegraphics[width=0.45\textwidth]{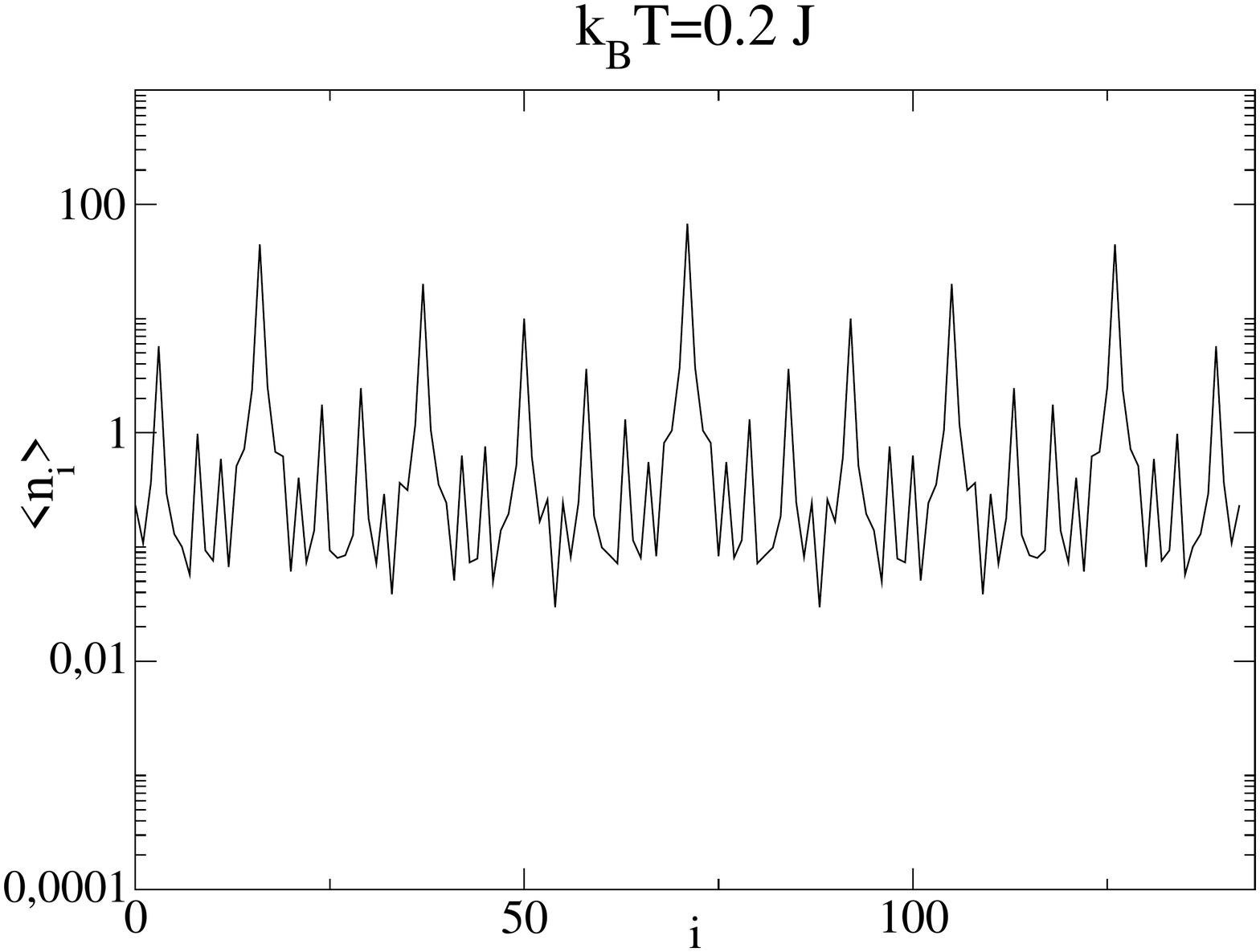}
\caption{Mean number of particles in each lattice obtained in the canonical ensemble for $M=144$ sites, $N=300$ particles, fixed value of the disorder strength $\Delta=2.5\,J$, calculated for various temperatures.
}
\label{pic_mean}
\end{figure}

\begin{figure}
\centering
\includegraphics[width=0.45\textwidth]{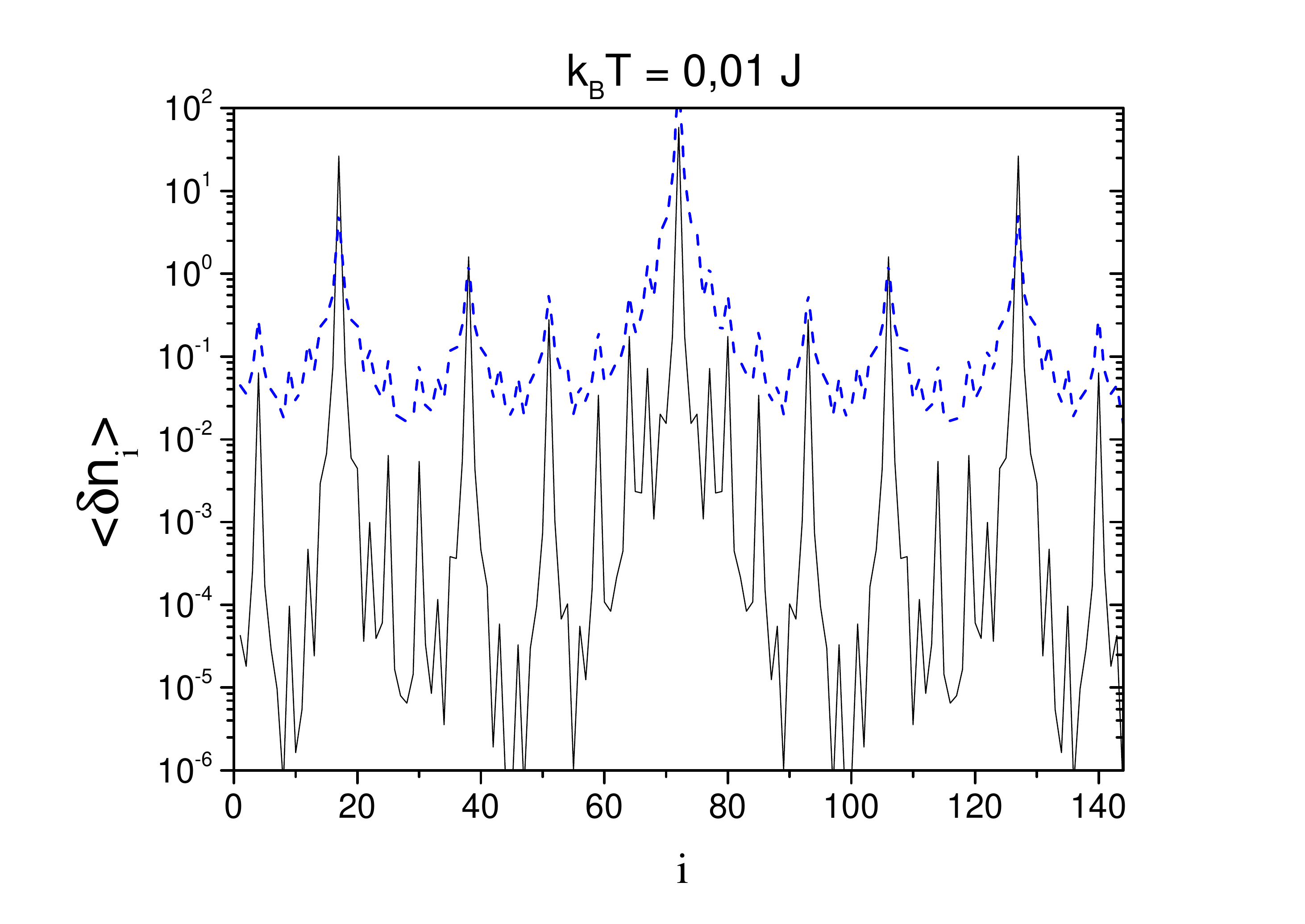}
\includegraphics[width=0.45\textwidth]{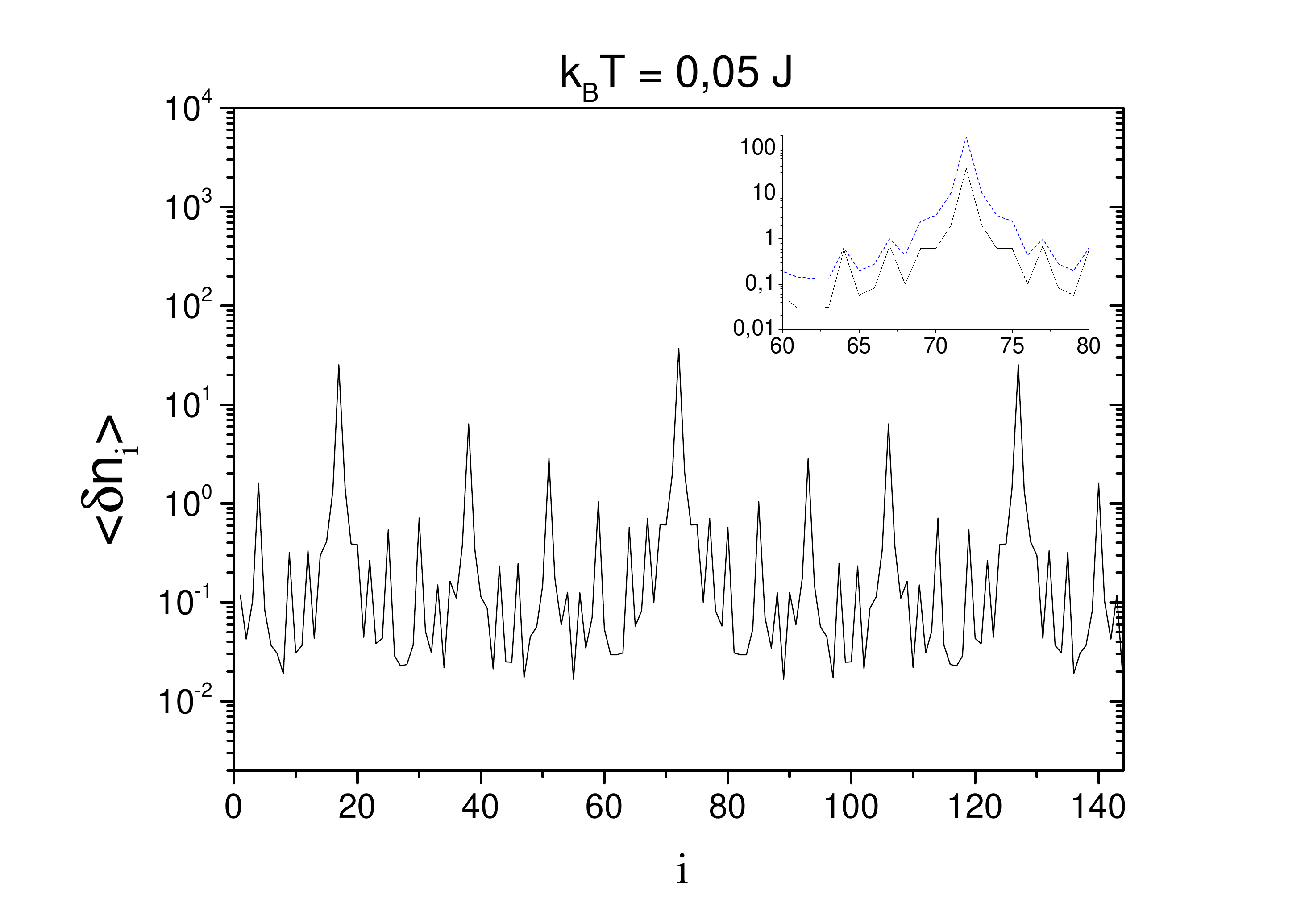}
\includegraphics[width=0.45\textwidth]{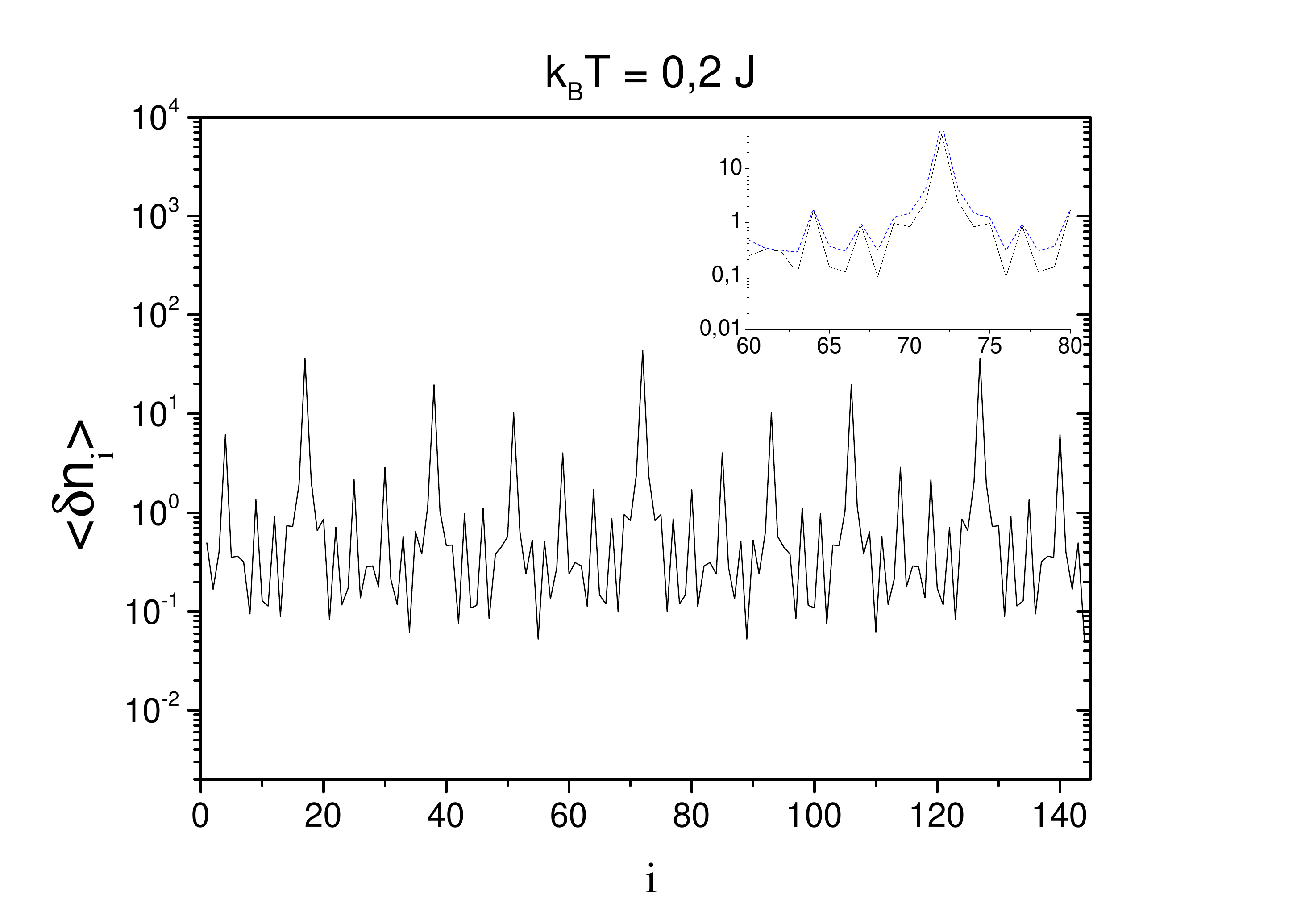}
\caption{Number fluctuations of particles in each lattice site obtained in the canonical ensemble for $M=144$ sites, $N=300$ particles, fixed value of the disorder strength $\Delta=2.5\,J$, calculated for various temperatures. The insets show comparison of exact results (black) with thermal approximation (blue, dashed) around the central peak.}
\label{pic_fl}
\end{figure}

In this section we analyze the mean and fluctuations of the number of particles in the wells of the optical lattice. In Figs.~\ref{pic_mean} and \ref{pic_fl} we show the mean occupation numbers and its fluctuations calculated for some sample parameters: $N=300$ particles, disorder amplitude $\Delta=2.5\,J$, and for various temperatures of the atomic gas. The peaks correspond to localized states that are centered at various lattice sites. At low temperatures the peaks are distributed symmetrically around the central peak, corresponding to the ground state. The remaining peaks result from the contribution of excited states. As the temperature increases, the number of populated states gradually grows and so does the number of peaks. Similar effect can be observed for fluctuations. It turns out that the behaviour of fluctuations can be qualitatively understood assuming thermal character of the fluctuations for each lattice site separately
\begin{equation}
\left\langle\delta^2 n_i\right\rangle = \left\langle n_i\right\rangle(\left\langle n_i\right\rangle+1),
\label{gcf}
\end{equation}
Similar result can be also obtained when considering the lattice as a set of separated potential wells, and describing the statistics of the single well within the grand-canonical ensemble in the equilibrium with the rest of the lattice sites, which can be treated as a reservoir~\cite{phd}. As a result we obtain formula~\eqref{gcf}. This approximation works particularly well at high temperatures. The comparison of exact results and the model is presented on Figure \ref{pic_fl}.

\subsection{Correlations between sites}
\begin{figure*}
\centering
\includegraphics[width=0.32\textwidth]{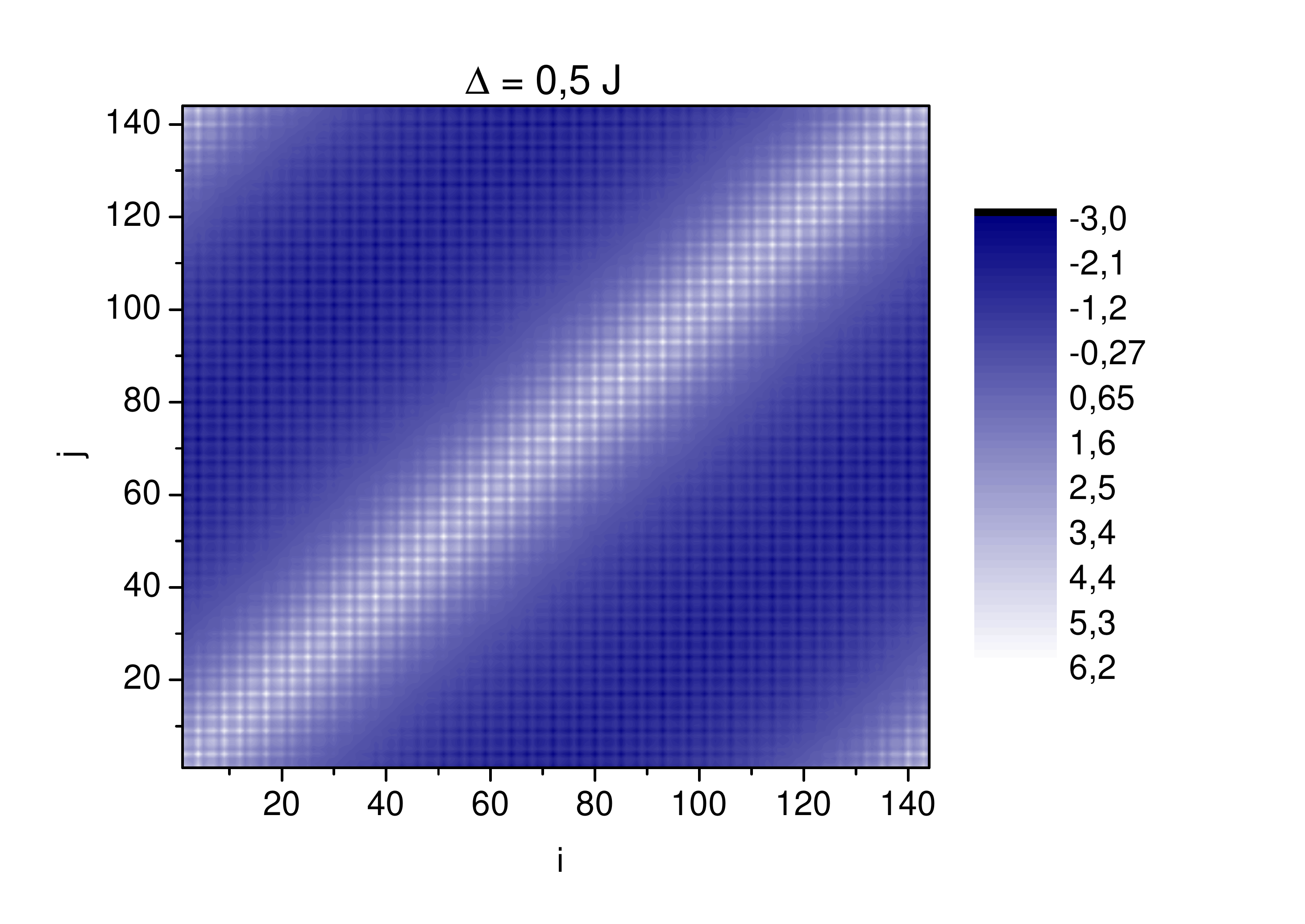}
\includegraphics[width=0.32\textwidth]{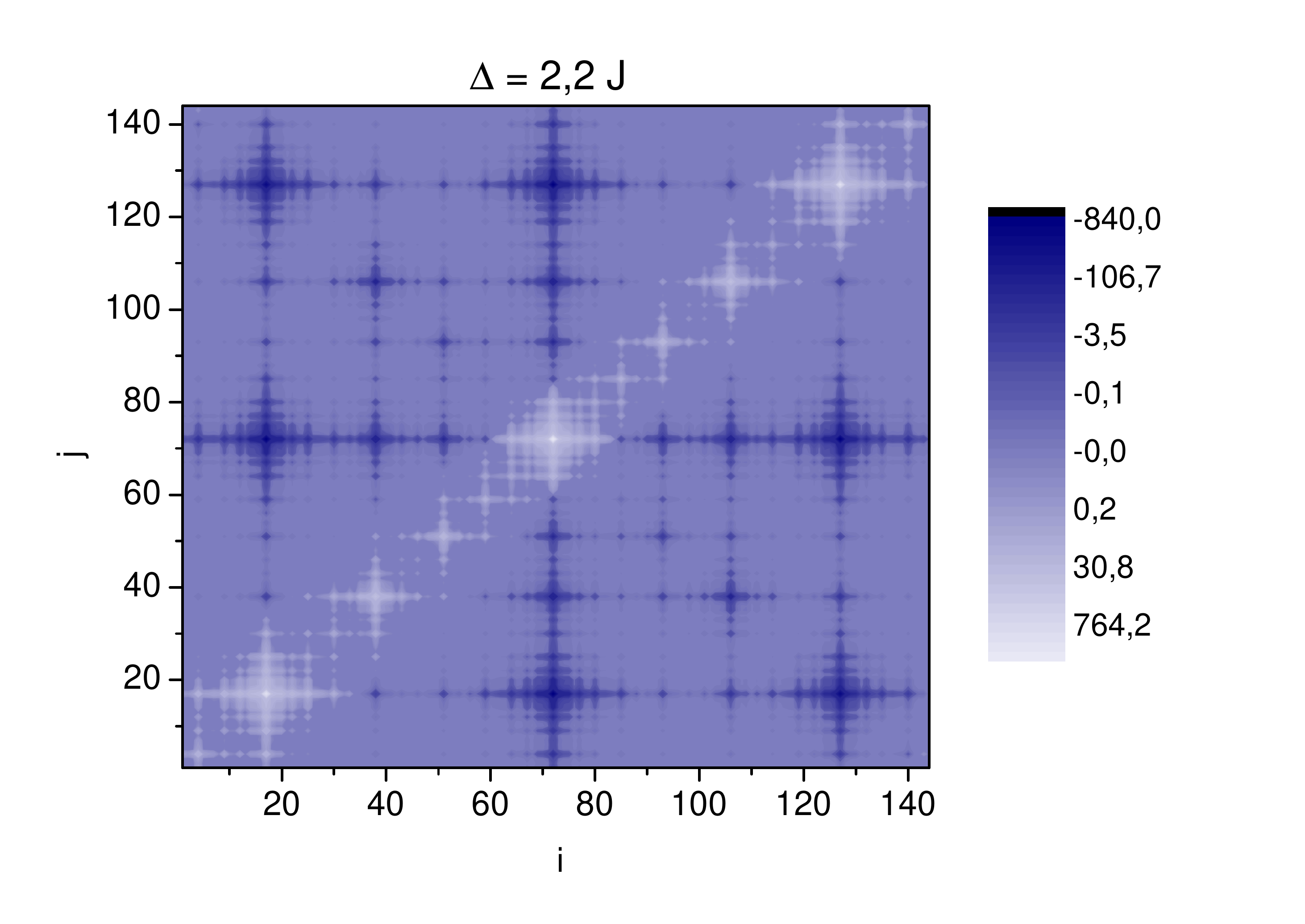}
\includegraphics[width=0.32\textwidth]{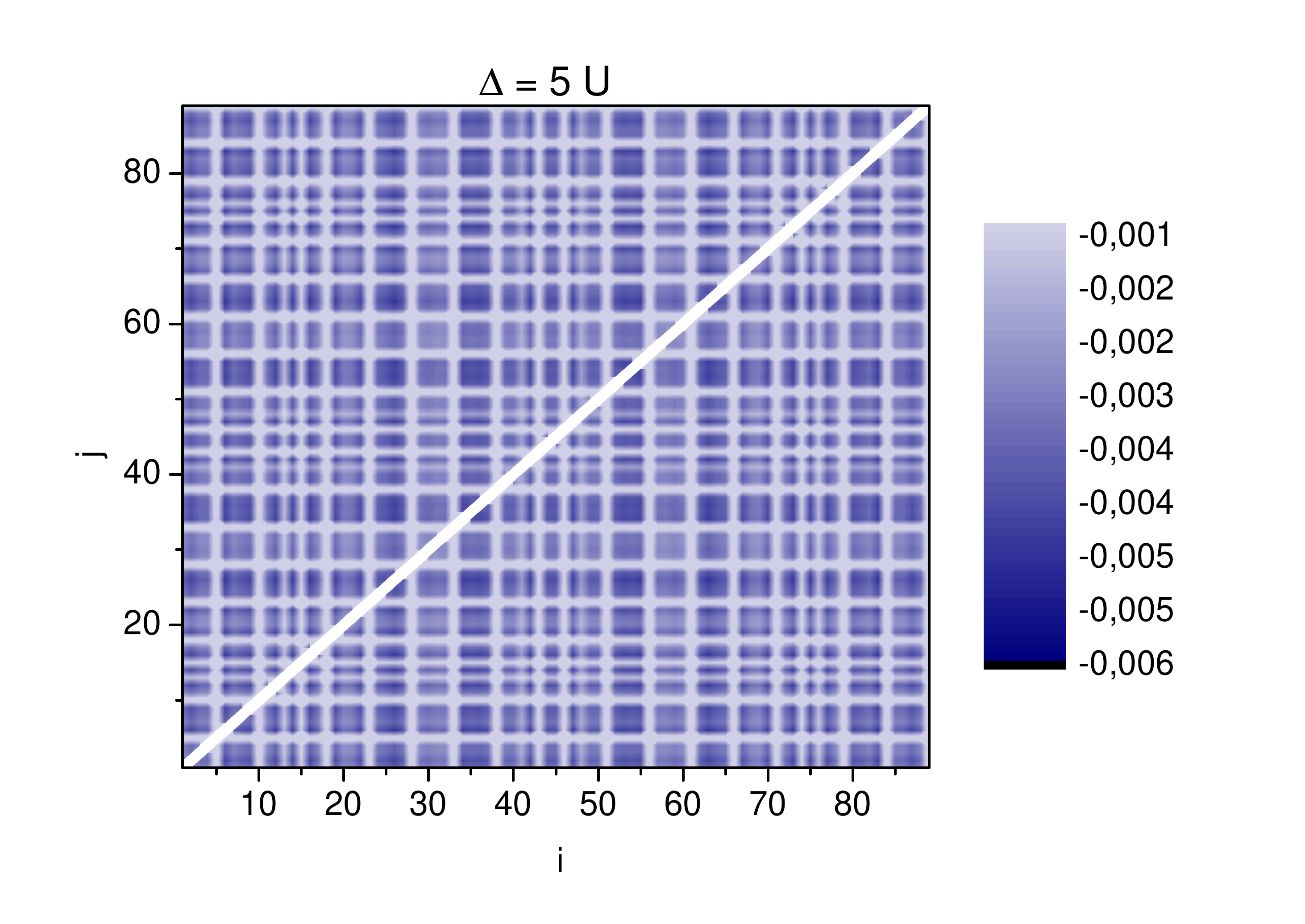}
\caption{Correlations between sites $\left\langle n_i n_j\right\rangle-\left\langle n_i\right\rangle\left\langle n_j\right\rangle$, for Aubry-Andre model with $k_B T = 0.05\,J$ and $\Delta=0.5\,J$ (left);  $k_B T = 0.05\,J$ and $\Delta=2.2\,J$ (middle); and for strongly interacting gas with $k_B T=0.5\,U$, $M=89$, $N=2M$ and $\Delta=5U$ (right).}
\label{corr1}
\end{figure*}

For completeness of the analysis we present the correlations $\left\langle n_i n_j\right\rangle-\left\langle n_i\right\rangle\left\langle n_j\right\rangle$ between the number of particles in different lattice sites, calculated in the canonical ensemble. The correlations for various values of disorder strength are shown in Fig.~\ref{corr1}. It appears that the correlations between strongly occupied sites are the largest. For large disorder the correlations are mainly negative, which results from the conservation of the total number of particles in the canonical ensemble: the more particles occupy certain localized state, the less are left for the other states. The positive values appear only between the lattice sites where the localization occurs. In contrast, at low disorder, the correlations are positive only between neighboring sites, which correspond to the diagonal and to the corners of the graph. This results from the fact that at low disorder all the excited states are spread along several sites.

\section{Strongly interacting gas in a quasi-periodic potential}

\subsection{Hamiltonian}

When the particles are strongly interacting, we may neglect the tunneling term in the Hamiltonian \eqref{hamiltonian} in comparison to the remaining terms. This yields
\begin{equation}
\label{HMott}
H=U\sum_k{n_k (n_k-1)}+\Delta \sum_k{n_k \cos(2 \pi \beta k)}.
\end{equation}
Now, there are only two energy scales given by $U$ and $\Delta$, and in the subsequent analysis we will express $\Delta$ in units of $U$. In the Hamiltonian \eqref{HMott} the different lattice sites are decoupled, thus the only correlation between sites is due to the conservation of the total number of particles. For such a system the partition function
\begin{equation}
Z(N,\beta)=\sum_{n_1}{}\ldots\sum_{n_M}{e^{-\beta\sum{\varepsilon_i n_i}-\beta U\sum{n_i(n_i-1)}/2}\delta_{\sum{n_i},N}}
\label{x}
\end{equation}
can be calculated exactly using a recurrence relations. We have developed a recurrence algorithm to calculate the partition function \eqref{x}, which can be derived by adding one lattice site in each step of the recurrence (see Appendix C for details).

\subsection{Statistics of the strongly interacting gas in the presence of disorder}

\begin{figure}
\centering
\includegraphics[width=0.4\textwidth]{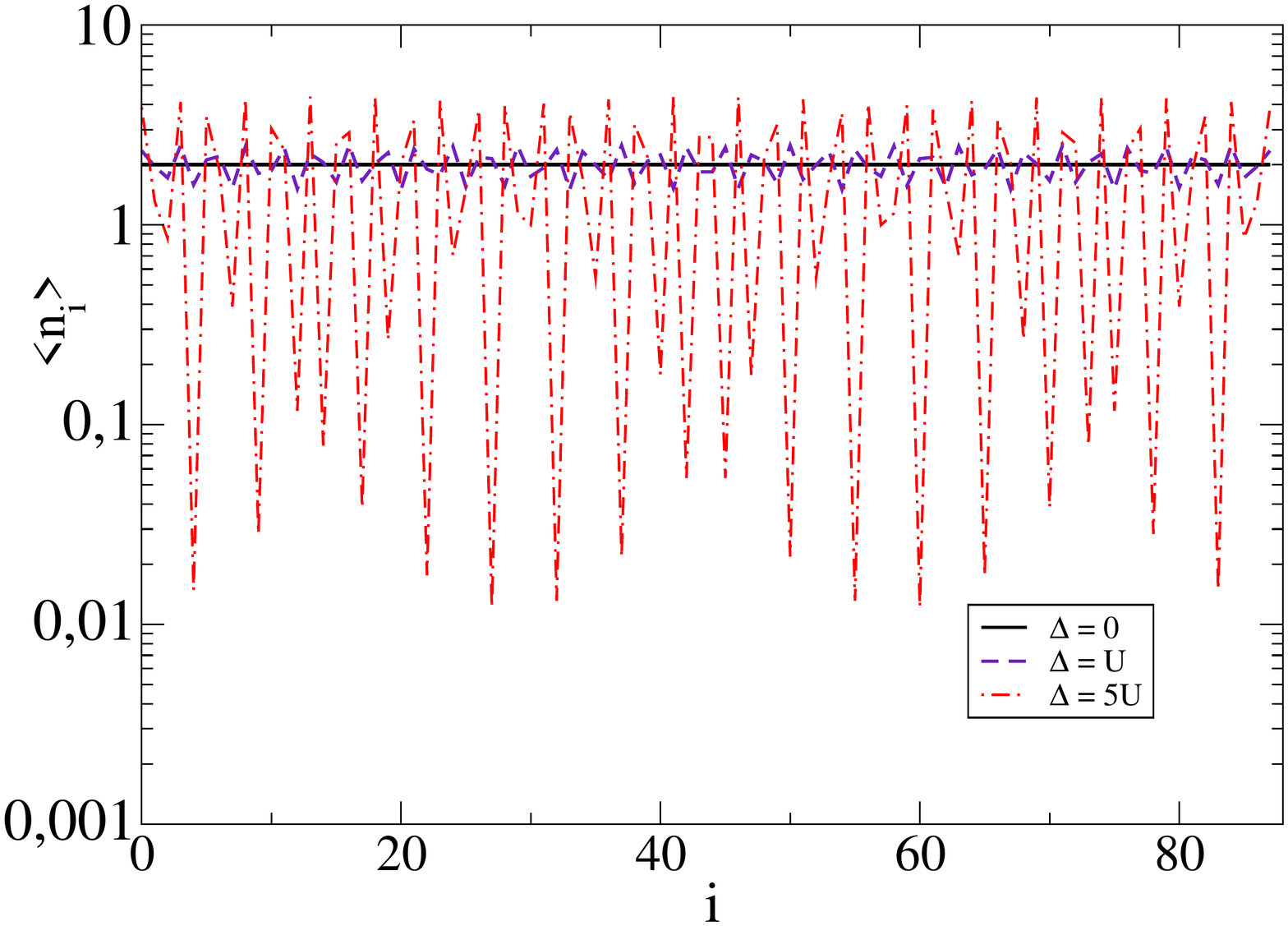}
\includegraphics[width=0.4\textwidth]{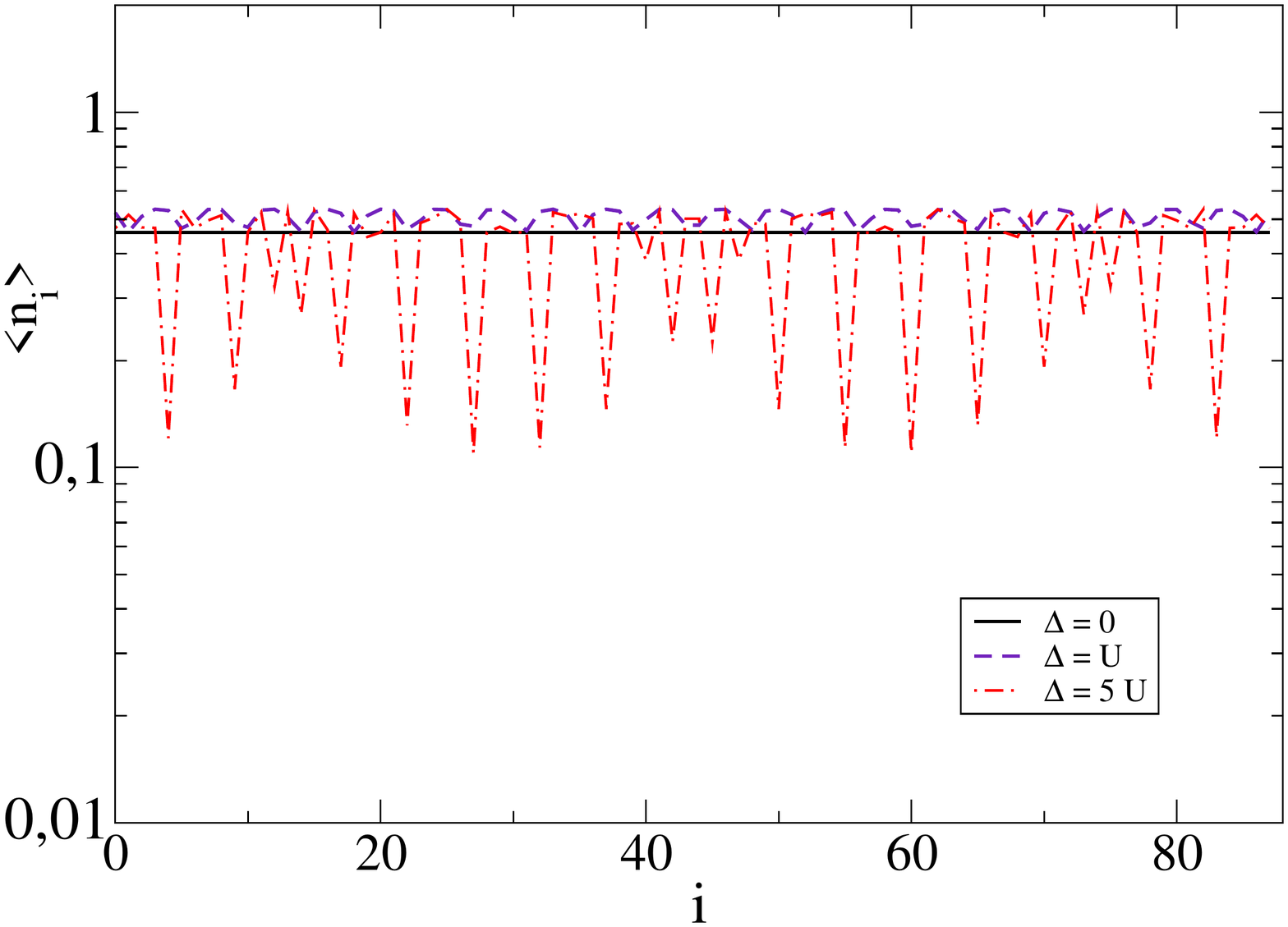}
\caption{Sample graph for mean occupation number (top) and its fluctuations (bottom) for $M=89$ sites, $N=2M$, $k_B T=0.5\,U$ and various $\Delta$.}
\label{mott}
\end{figure}

We have analyzed the mean particle number and fluctuations in the wells in the case of strongly interacting gas in quasi periodic potentials. As the tunneling process is neglected, the localization is not presents and the only effects influencing the mean and fluctuations results form the variation of the chemical potential at different lattice wells. This statement is confirmed by the analysis of numerical results shown in Fig.~~\ref{mott}. We have performed numerical calculations for a moderate-size system containing $M=89$ sites and $N=2 M$ particles. We observe that both mean and fluctuations vary stronger from site to site as the amplitude of disorder increases. The correlations between occupation numbers at different sites for some example value od $\Delta$ are presented in Fig.~\ref{corr1}. As the sites are uncoupled in the Hamiltonian \eqref{HMott}, the correlations result only from the constraint on the total number of particles, and they are strongest for sites with highest occupation numbers.

\section{Probing the statistical properties of the system with light scattering}

We consider the possibility of distinguishing between different many-body phases of ultracold atoms in disordered potentials. This can be done, for instance, by measuring the properties of the correlation functions. One of the possible tools that can bring information about the correlation function is the measurement of the properties of light scattered on ultracold atoms. Previously, the atom-light interactions were suggested as a method to detect Bose-Einstein condensation in an ultracold gas~\cite{Lewenstein1993}, BCS transition in ultracold fermions \cite{Zhang1999}, statistics
of ultracold atomic gases~\cite{Idziaszek}, or distinguishing between quantum phases of ultracold atom in optical lattices \cite{Mekhov2007,Lakomy,Menotti2010,Douglas2011}.

Let us consider the gas of ultracold atoms in an external potential interacting with a weak and far-detuned laser with frequency $\omega_L$. Treating the atoms as two-level systems, it is possible to adiabatically eliminate the excited state and obtain the effective hamiltonian. In this way we can calculate the mean number of scattered photons with wave vector $\mathbf{k}$ and polarization $\epsilon$ per unit time per solid angle~\cite{Lakomy}
 \begin{equation}
\frac{\partial^2N}{\partial t\partial\Omega} = \frac{\Omega^2 c_k ^2}{\delta_L^2}\frac{\pi}{2} F(\mathbf{q}),
 \label{scattering}
 \end{equation}
where $\Omega =E_l \mathbf{\epsilon}_L \mathbf{d}/\hbar$ is the Rabi frequency, $\delta_L$ is the detuning of the laser, $c_k=g_k \mathbf{\epsilon}_k \mathbf{d}/\hbar$, $\mathbf{q}=\mathbf{k}-\mathbf{k_L}$, $\left|k\right|=\left|k_L\right|$ (elastic scattering), $E_l$ stands for the electric field of the laser with polarization $\mathbf{\epsilon}_L$, $\mathbf{d}$ is the atomic dipole moment, $g_k$ is the coupling constant  and
\begin{equation}
F(\mathbf{q})=\int{d^3 x}\int{d^3 y\,e^{i\mathbf{q}(\mathbf{x}-\mathbf{y})}\left\langle\psi^\dag (\mathbf{x}) \psi(\mathbf{x}) \psi^\dag (\mathbf{y}) \psi(\mathbf{y})\right\rangle}
\end{equation}
carries the information about the statistics of the system. Function $F(\mathbf{q})$ is defined as the Fourier transform of the second correlation function. It is equivalent to the static structure factor~\cite{Lakomy} and in the rest of the work we will refer to $F(\mathbf{q})$ as the static structure factor.

\begin{figure}
\centering
\includegraphics[width=0.33\textwidth]{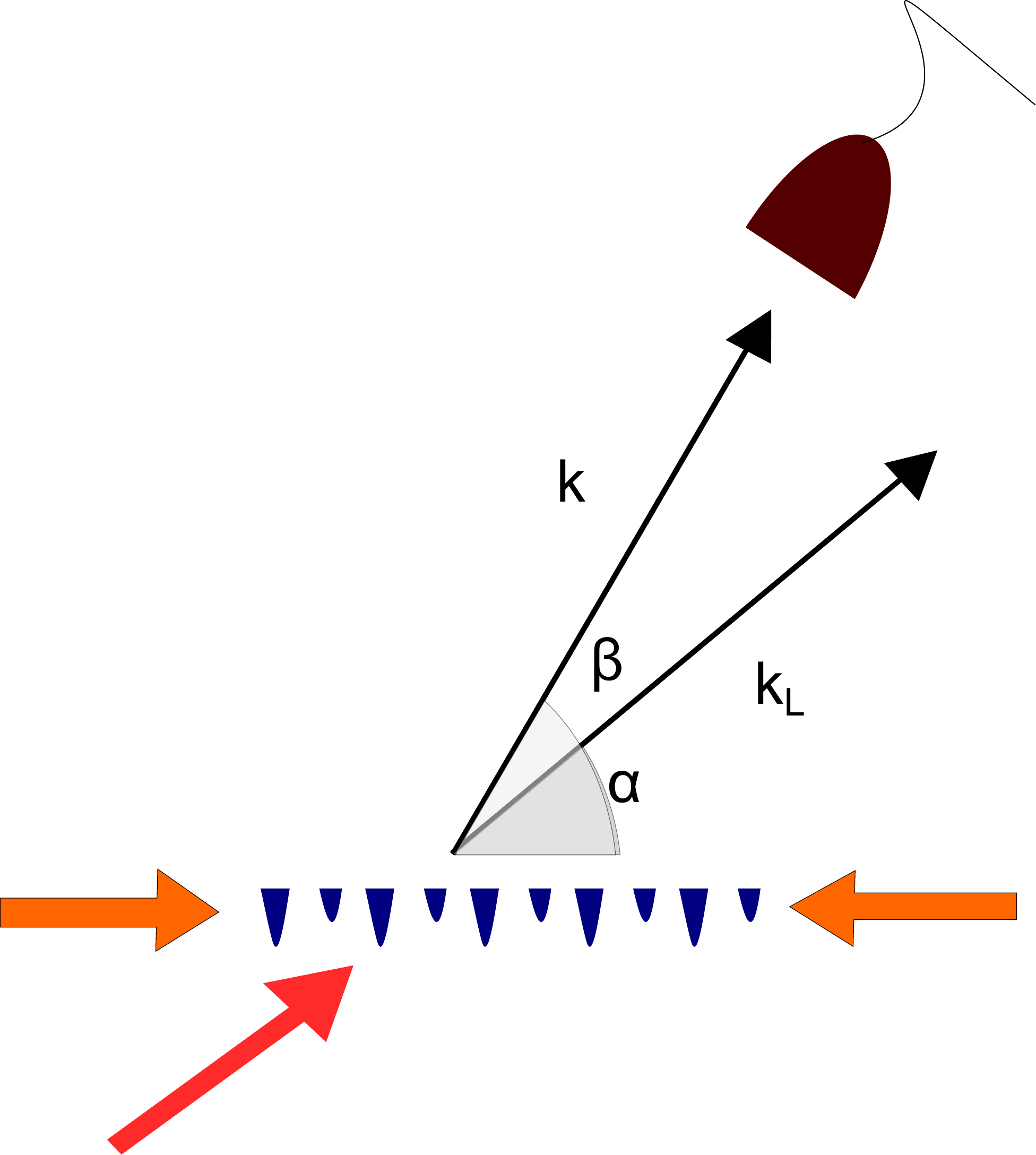}
\caption{Experimental setup for light scattering. The optical lattice is illuminated by the laser set at angle $\alpha$. The detector is set at angle $\beta$.}
\label{setup}
\end{figure}

\subsection{Light scattering from bosons in an optical superlattice}
\begin{figure*}
\includegraphics[width=0.45\textwidth]{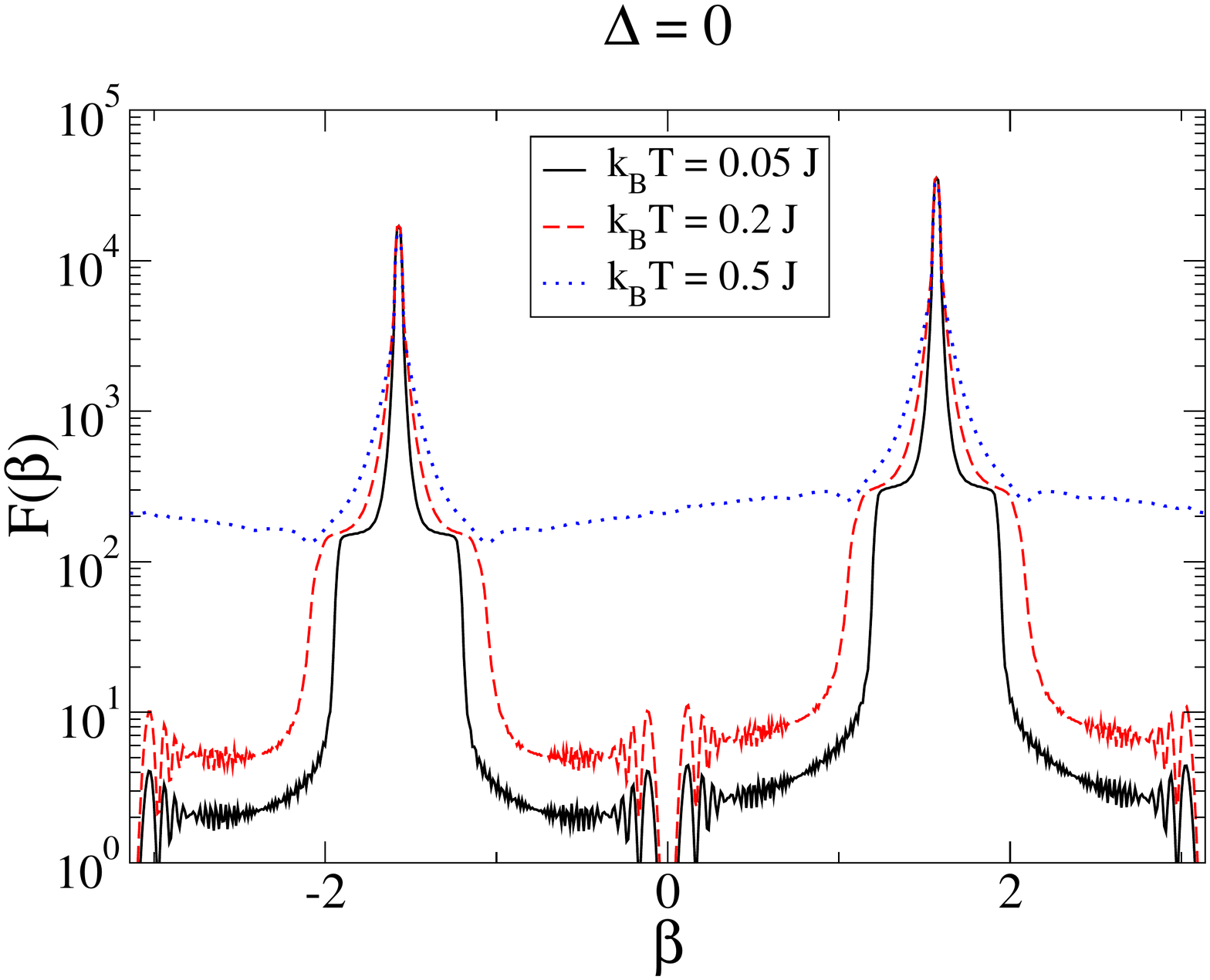}
\includegraphics[width=0.45\textwidth]{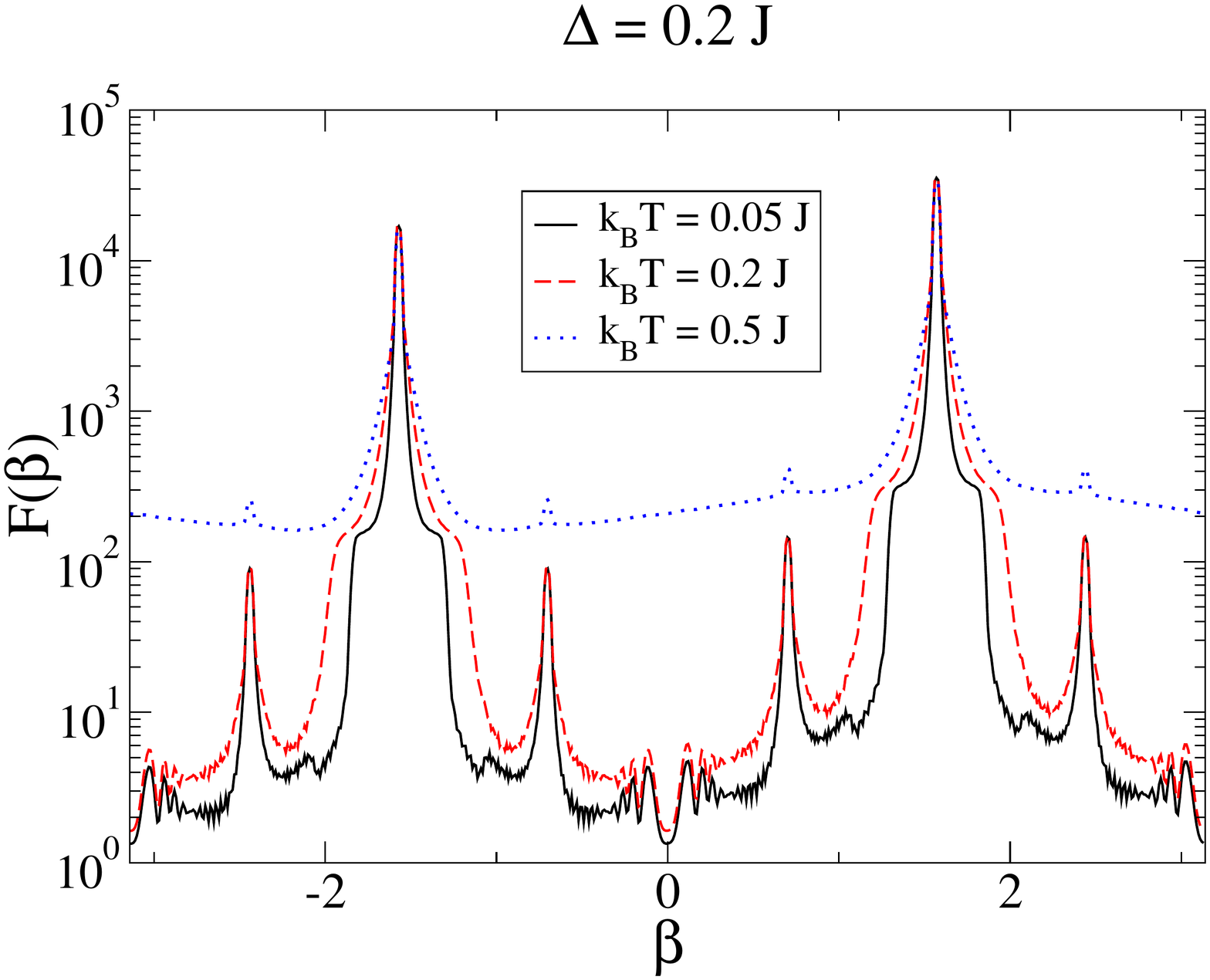}\\
\includegraphics[width=0.45\textwidth]{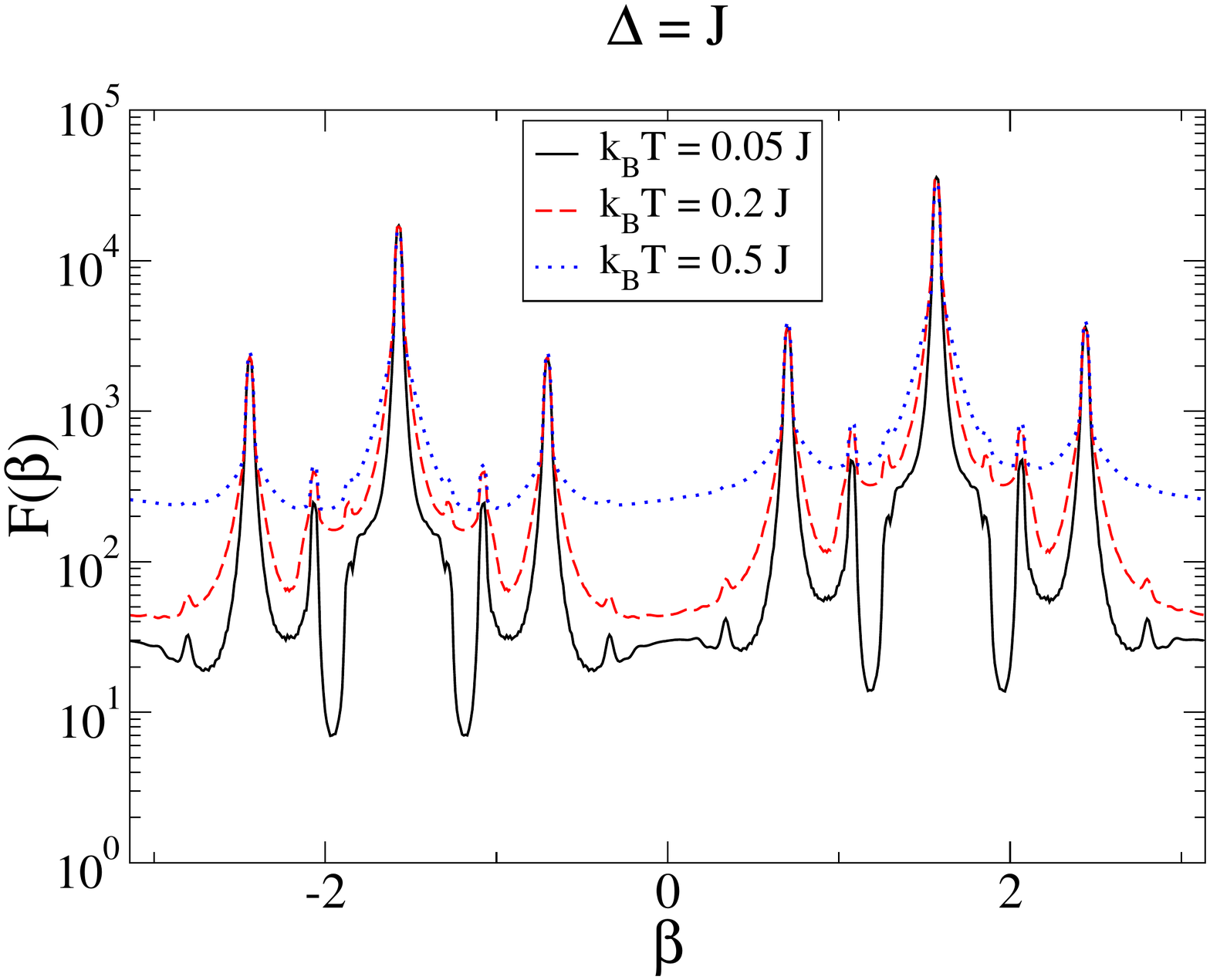}
\includegraphics[width=0.45\textwidth]{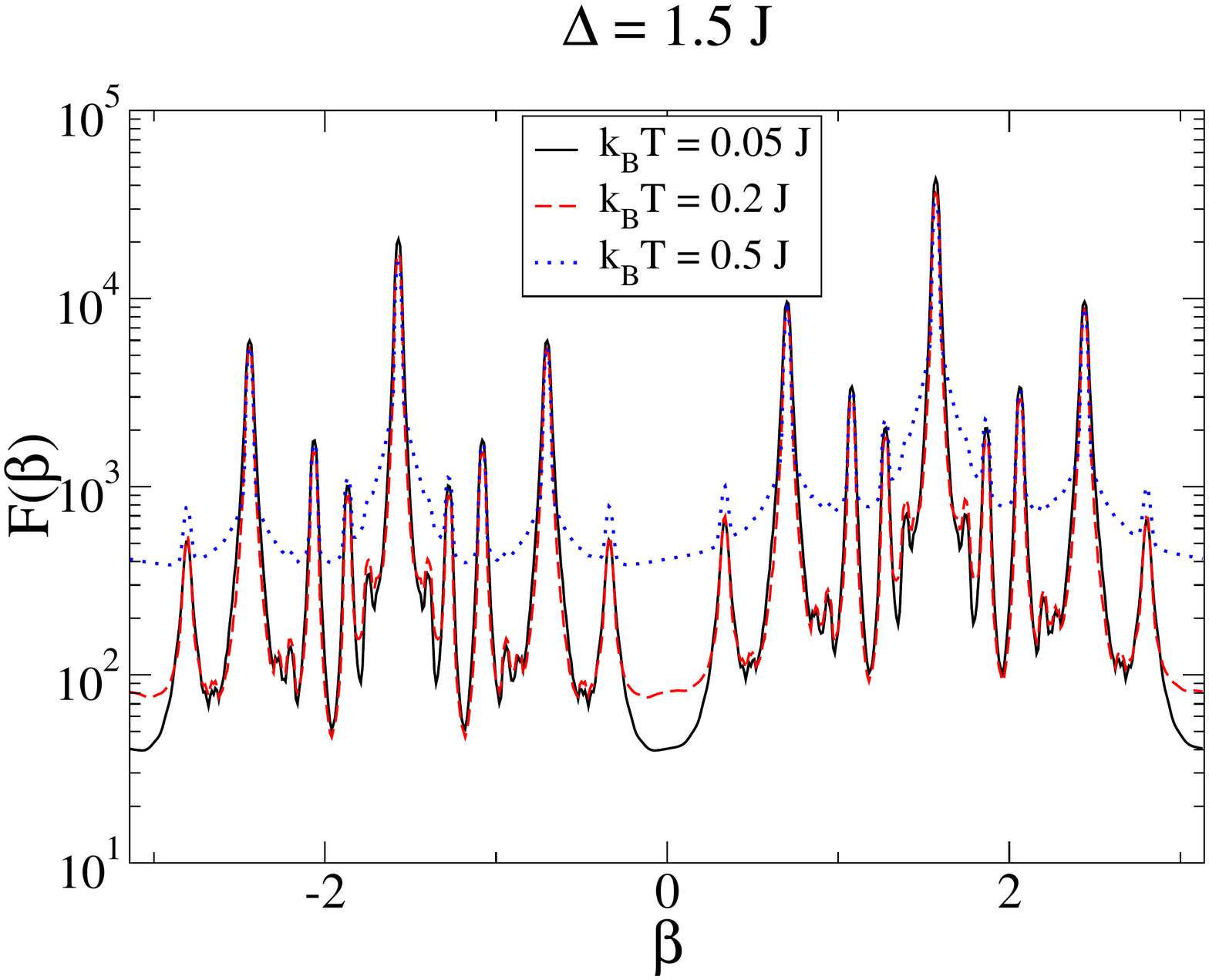}\\
\includegraphics[width=0.45\textwidth]{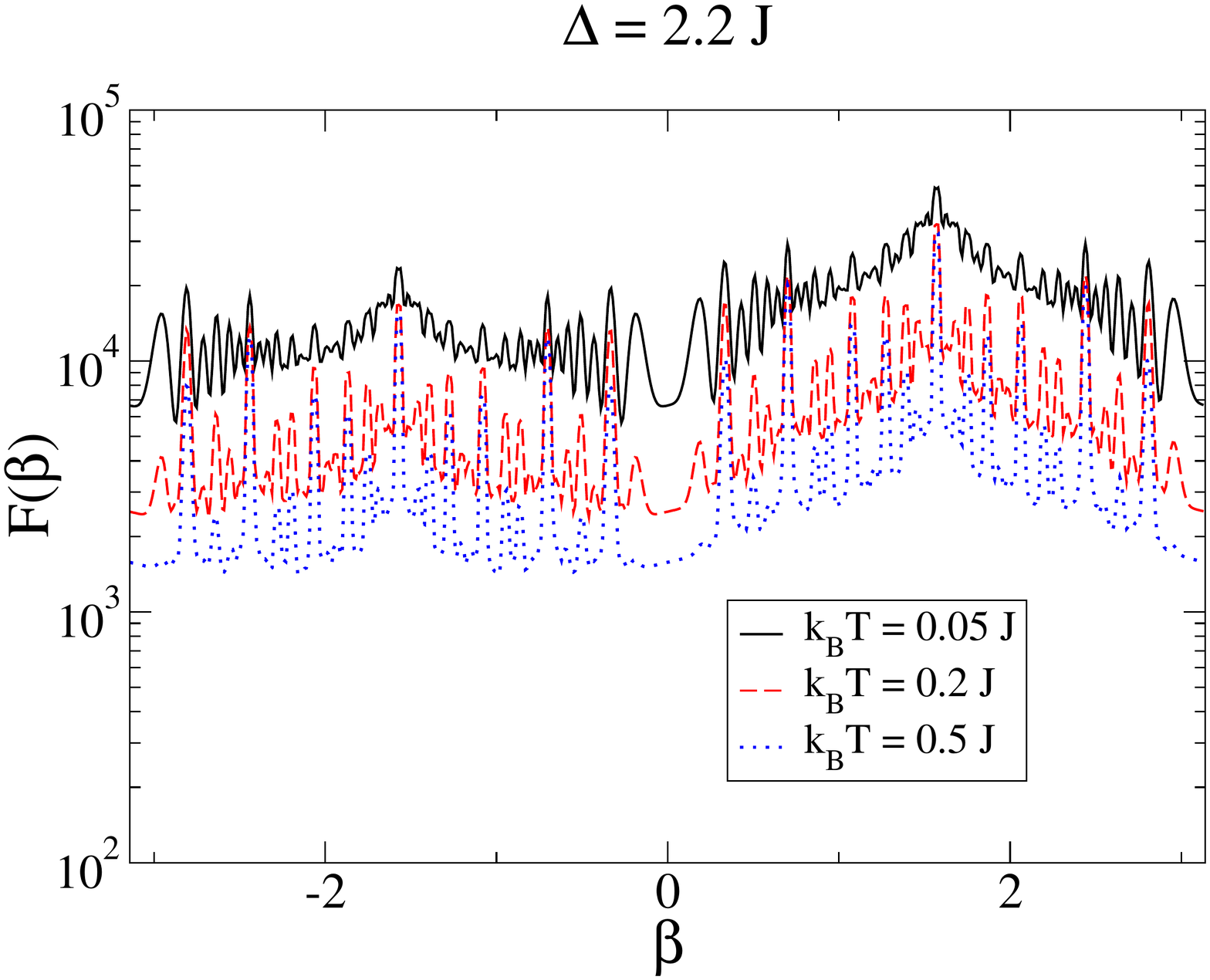}
\includegraphics[width=0.45\textwidth]{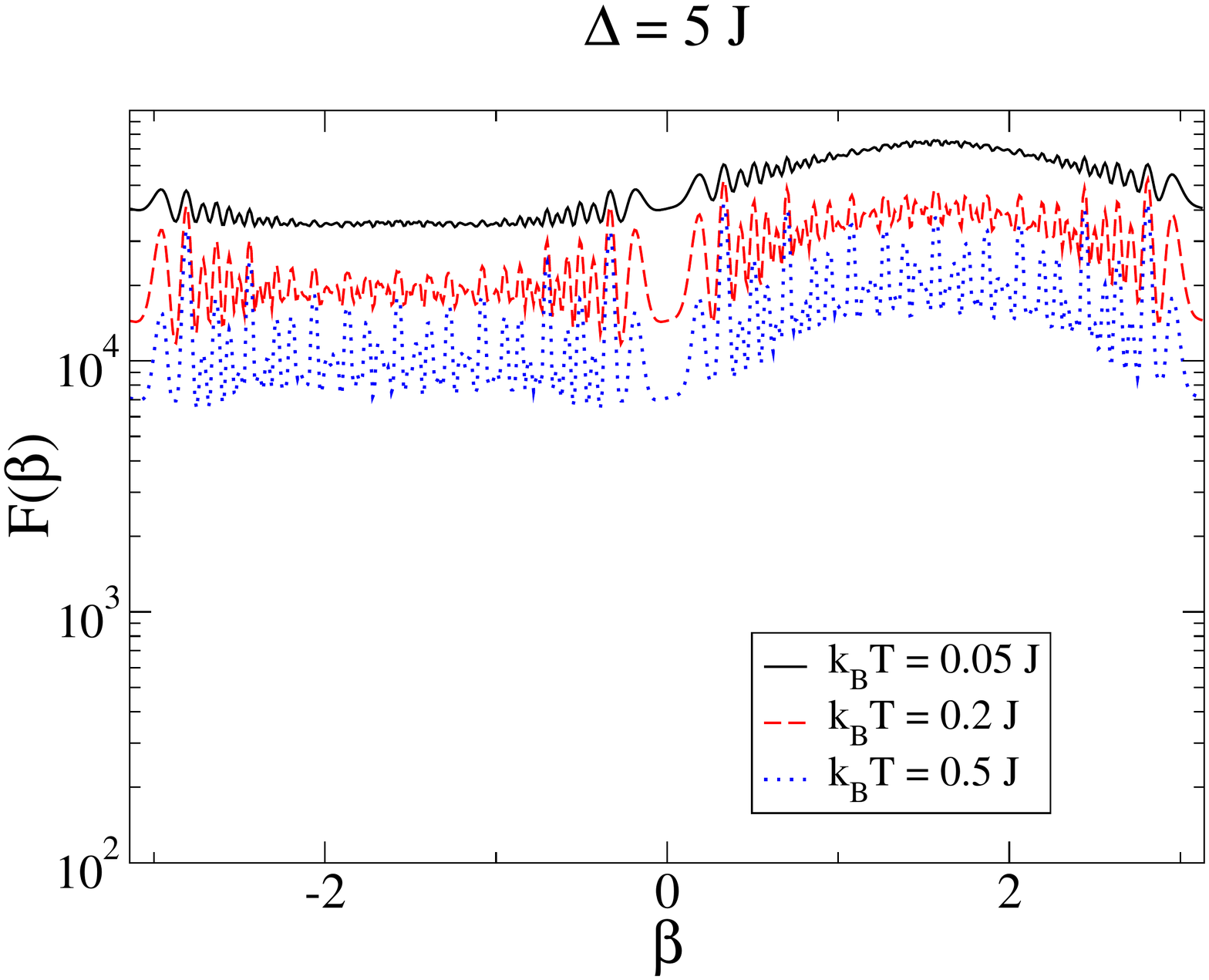}
\caption{The averaged spectrum of scattered photons for Aubry-Andre model for different values of $\Delta$ and $T$. The probing beam is set at the angle $\alpha=\pi/2$. For $\Delta<2\,J$ the system is in the superfluid state while for larger $\Delta$ it is in the localized regime.}
\label{spektra}
\end{figure*}
We now show how to extract the information on correlations from the intensity of light scattered at different angles. For a single atom, one can show that $F(\mathbf{q})=1$ \cite{Idziaszek}, so $F$ represents the difference between scattering from one atom and from the many-body system. In the following, we will focus solely on the properties of the structure factor. By expanding the field operators into Wannier states, we get an equivalent formula for $F$:
\begin{equation}
F(\mathbf{q}) =\sum_{n,n',m,m'}{\left\langle n\left|e^{i\mathbf{qr}}\right|n'\right\rangle\left\langle m\left|e^{-i\mathbf{qr}}\right|m'\right\rangle\left\langle g_n^\dag g_{n'}g_m^\dag g_{m'}\right\rangle}
\label{ffunc}
\end{equation}
The matrix elements $\left\langle n\left|e^{i\mathbf{qr}}\right|n'\right\rangle$ are calculated between Wannier states localized in sites $n$ and $n'$. We will consider the deep lattice regime where Wannier states are strongly localized and hence the terms with $n\neq n'$ are negligible. This approximation is valid when the lattice potential depth is of the order of several recoil energies. In this regime we may also use gaussian approximation of the Wannier states. Formula (\ref{ffunc}) simplifies to
\begin{eqnarray}
F(\mathbf{q}) =\sum_{n,m}{\left\langle n\left|e^{i\mathbf{qr}}\right|n\right\rangle\left\langle m\left|e^{-i\mathbf{qr}}\right|m\right\rangle\left\langle n_n n_m\right\rangle}=\nonumber\\=\sum_{n,m}{\left|f_0(\mathbf{q})\right|^2 e^{i\mathbf{q}(\mathbf{r}_n-\mathbf{r}_m)}}\left\langle n_n n_m\right\rangle,
\label{ffunc2}
\end{eqnarray}
where
\begin{equation}
f_0(\mathbf{q})=\int{d^2r\,\left|w_0(\mathbf{r})\right|^2e^{i\mathbf{qr}}}.
\end{equation}

The term $\mathbf{q}(\mathbf{r}_n-\mathbf{r}_m)$ may be rewritten as $\gamma(n-m)$, where
\begin{equation}
\gamma = \pi \frac{\lambda}{\lambda_L}(\cos\beta-\cos\alpha),
\end{equation}
$\lambda$ is the wavelength of the laser forming the primary optical lattice and $\lambda_L$ is the wavelength of the probing laser. Angle $\beta$ is the angle at which the detector is set and $\alpha$ is the angle of the probing laser (see Figure \ref{setup}).

It is instructive to split $F(\mathbf{q})$ into two parts $F_{class}$ and $F_{quant}$ \cite{Mekhov2007}, where
\begin{equation}
F_{class}(\mathbf{q})=\left|f_0(q)\right|^2\left|\sum_m{e^{i\mathbf{qr_m}}\left\langle n_m\right\rangle}\right|^2
\end{equation}
represents the so called classical component of the scattered light. It is obtained by calculating the average $| \langle a_{\mbf{k}\lambda} \rangle|^2$, which is proportional to the amplitude of the electric field square. The difference between the total function $F(\mbf{q})$ and the classical part defines the quantum component~\cite{Mekhov2007}
\begin{equation}
F_{quant}(\mathbf{q})=\left|f_0(q)\right|^2\sum_{n,m}{e^{iq(r_n-r_m)}\left(\left\langle n_n n_m\right\rangle-\left\langle n_n\right\rangle\left\langle n_m\right\rangle\right)}
\end{equation}
It gives information about quantum statistical effects in the system. Splitting $F(\mathbf{q})$ into these two parts is particularly useful when comparing Mott insulator and superfluid phases, as both of them are homogenous so they differ only in the quantum component~\cite{Mekhov2007,Lakomy}. Here this will not be the case, as the system is inhomogeneous and already the classical components of various quantum phases are different.

For the homogenous phase with density $n$, the classical part of $F$ can be expressed as
\begin{equation}
F_{class}(\mathbf{q})=\left|f_0(q)\right|^2 n^2 \frac{\sin^2\left(M\pi\lambda/2\lambda_L (\cos\beta-\cos\alpha)\right)}{\sin^2\left(\pi\lambda/2\lambda_L (\cos\beta-\cos\alpha)\right)}
\end{equation}
which gives us intuition that as $M$ or $\lambda/\lambda_L$ increases, $F(\mathbf{q})$ should oscillate faster. This quantity has already been measured in experiment for a two-dimensional Mott insulator \cite{Bloch2011}.

\subsection{Scattering from localized and delocalized phases}

We now use the method described above to analyze the possibility to distinguish localized and delocalized phases in Aubry-Andre model. We will use the results obtained in the canonical ensemble and presented in the previous chapter. There are many parameters which can be varied in calculations and in experiment: the number of particles $N$, number of lattice sites $M$, temperature $T$, primary lattice depth $s_1$, probe laser wavelength $\lambda$ and the angle at which the detector is set $\alpha$. We set $N=300$, $M=144$, $s_1=5$, $\lambda=\lambda_l$ (the lattice laser wavelength), and $\alpha =\pi/2$ and examine how the spectrum of scattered photons changes with growing $\Delta$ and temperature.

As shown on Figure \ref{spektra}, the growth of $\Delta$ causes additional interference peaks to emerge. This results from the influence of the second lattice, which generates additional momenta $k_2$, $k_1-k_2$ etc. in the system. Similar observation was made in \cite{Guarrera2007}, where the impact of the second lattice on the noise correlations was studied experimentally. As $\Delta$ crosses the transition point, due to incommensurability of the lattices, the angular distribution flattens as the number of interference peaks goes to infinity. As a result we are able to detect localization for high $\Delta$, as well as observe the growing impact of the secondary lattice for low disorder.

High temperature rises the number of excited particles and disturbs the angular distribution of photons in two ways. Below the transition point $\Delta = 2\,J$ it reduces the visibility of the interference peaks. For higher $\Delta$ the presence of several localized states produces the interference peaks in the distribution which would not be present at $T=0$.

\subsection{Scattering from strongly interacting gas}
\begin{figure}
\centering
\includegraphics[width=0.45\textwidth]{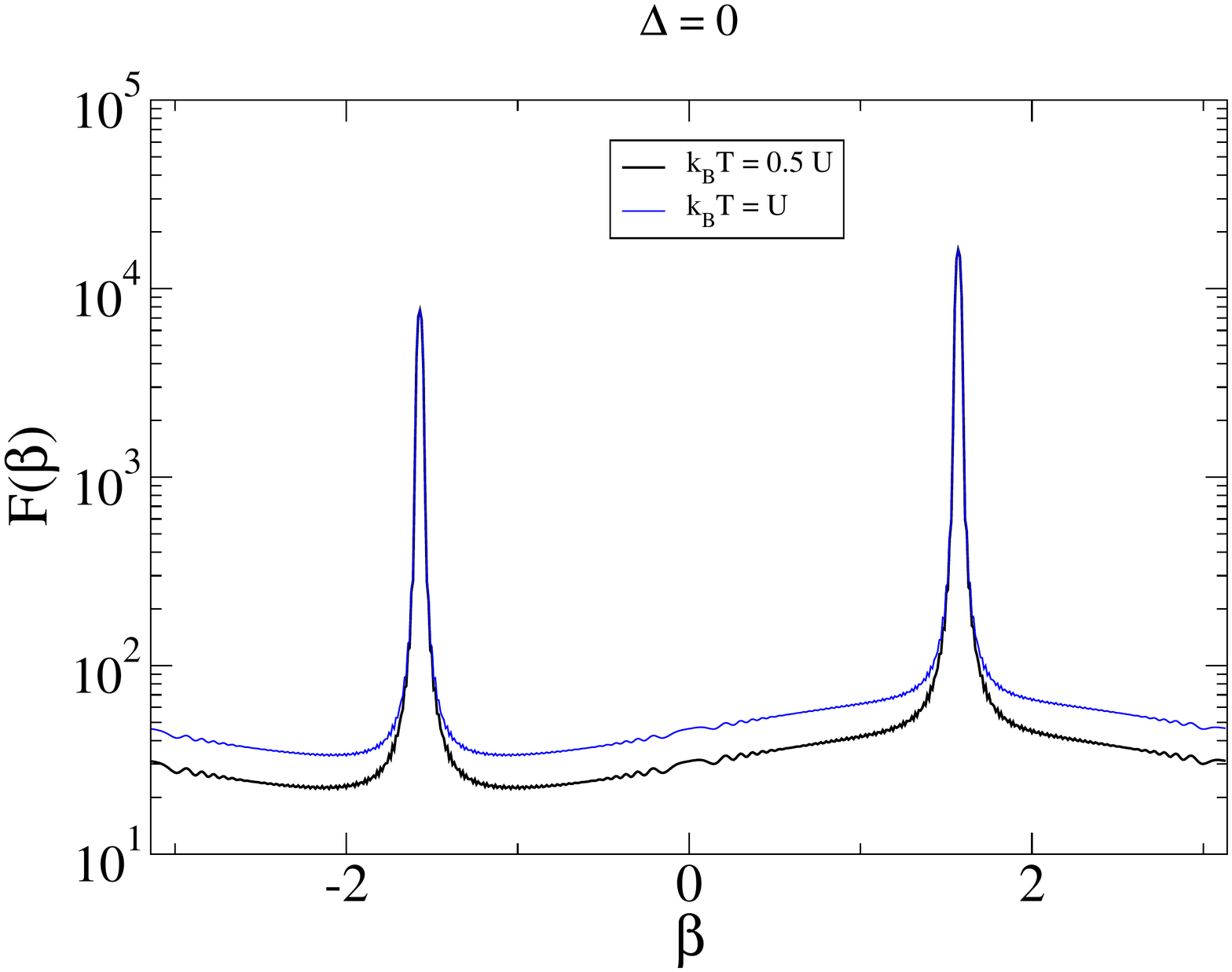}
\includegraphics[width=0.45\textwidth]{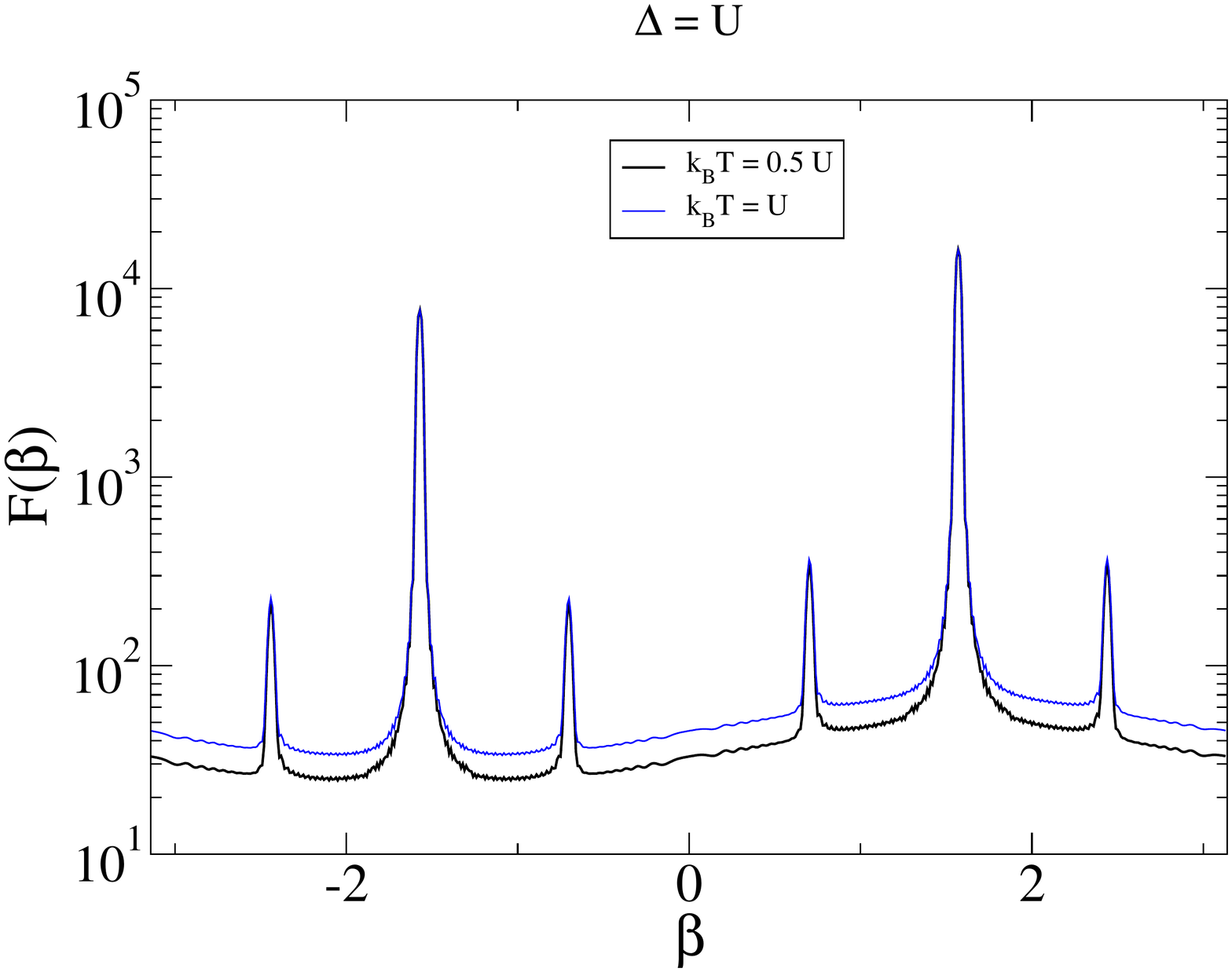}
\includegraphics[width=0.45\textwidth]{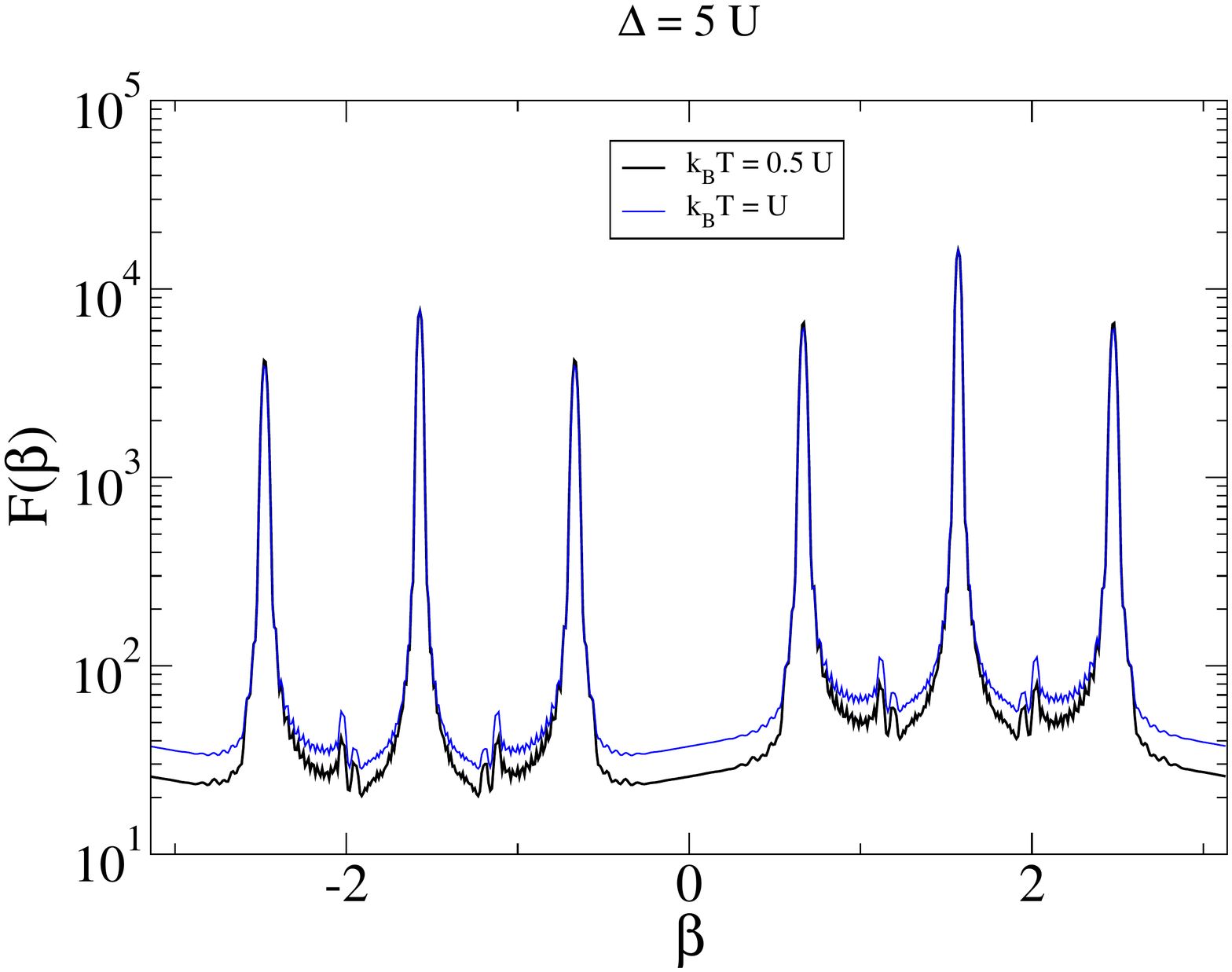}
\caption{The angular distribution of scattered photons for the strongly interacting gas for different values of $\Delta$ and $T$. The probing beam is set at the angle $\alpha=\pi/2$.}
\label{mottspektra}
\end{figure}

We now analyze the scattered spectrum for the strongly interacting gas, keeping the parameters of the bichromatic lattice unchanged and setting $M=89$ and $N=2M$. Sample pictures are shown on Figure \ref{mottspektra}. In the absence of disorder $\Delta=0$, the ultracold Bose gas forms a Mott insulator, and in such a case it was predicted that the angular distribution of scattered photons should exhibit the pattern of interference fringes, with a set of minima where there are no scattered photons \cite{Mekhov2007}. This is due to the contribution from the classical component, while the quantum part is zero due to the absence of the correlations. In contrast to the previous works, in our calculations we include the effects of the correlations between wells due to the constraint on the total number of particles in the canonical ensemble. This gives rise to the nonzero quantum component, and as the result the minima are no longer present in the angular distribution.

For growing disorder we again observe the appearance of new peaks in the spectrum. In this case they reflect the fact that the additional lattice of different period was added and the gas has a new density profile. However, there is no localization, so there is no qualitative change for growing $\Delta$. The angular distribution will flatten only in the limit $\Delta\rightarrow\infty$. This means that in principle it should be possible to distinguish the strongly interacting phase from the localized phase which may be useful in examining the role of interactions in disordered systems.

\section{Conclusions}
In conclusion, we studied the statistical properties of a Bose gas confined in a bichromatic optical lattice at finite temperature. We considered two limits, when the Hamiltonian can be diagonalized exactly: the ideal gas, when there is no interactions, and strongly interacting gas, when one can neglect the inter-well tunneling. We analized the mean, fluctuations and correlations between lattice sites occupation numbers for the Bose-condensed phase, localized phase and strongly interacting phase. We have shown that some important information about the structure factor can be extracted using light scattering, which makes possible to distinguish different phases and explore the phase diagram experimentally.
\section{Acknowledgements}
The authors would like to thank Micha\l$\,$Krych for carefully reading the manuscript. This work was supported by the Foundation for Polish Science International PhD Projects Programme co-financed by the EU European Regional Development Fund and by the National Center for Science grant number DEC-2011/01/B/ST2/02030.

\appendix
\section{Calculating the statistical quantities for an ideal gas}
In this appendix we derive a recurrence formula allowing for calculation of the partition function of the noninteracting gas exactly~\cite{Weiss97}. Let us denote the probability of finding exactly $n$ particles in a given state $\mu$ as $p_\mu (n)$. It may be calculated as $p^\geq _\mu (n)-p^\geq _\mu (n+1)$ which represent the probabilities of finding at least $n$ (or $n+1$, respectively) particles in an eigenstate $\mu$. Probability $p^\geq _\mu (n)$ is given by the formula
\begin{equation}
p^\geq _\mu (n)=\frac{1}{Z(\beta,N)}\sum_{n_1=0}^\infty{}\ldots \sum_{n_\infty=0}^\infty{}e^{-\beta \sum_\nu{n_\nu \epsilon_\nu}}\delta_{(\sum_i{n_i},N)}.
\end{equation}
By changing the summation index $\tilde{n}_\mu=n_\mu-n$ we obtain exactly a formula for $Z(\beta,N-n)$ with prefactor $e^{-n\beta\epsilon_\mu}/Z(\beta,N)$. Therefore,
\begin{equation}
p_\mu (n)= e^{-n\beta\epsilon_\mu}\frac{Z(\beta,N-n)}{Z(\beta,N)}-e^{-(n+1)\beta\epsilon_\mu}\frac{Z(\beta,N-n-1)}{Z(\beta,N)}
\label{oneprob}
\end{equation}
With this result we can easily calculate the mean occupation number of state $\mu$, directly from its definition $\left\langle n_\mu \right\rangle=\sum_{n=1}^N{n\, p_\mu(n)}$ and using \eqref{oneprob}, obtaining
\begin{equation}
\left\langle n_\mu \right\rangle=\frac{1}{Z_N}\sum_{n=1}^N{e^{-\beta n \epsilon_\mu}Z_{N-n}}.
\end{equation}
Summing  $\langle n_\mu \rangle$ over all states, we obtain the desired formula for the partition function
\begin{equation}
Z(\beta,N)=\sum_{n=1}^N{}\sum_\nu{e^{-n\beta\epsilon_\nu}Z(\beta,N-n)}.
\label{recurr}
\end{equation}
For calculating the fluctuations of the occupation number for a single site, as well as the correlations of the occupation numbers between different sites, we need to express  $\left\langle n_k n_l \right\rangle$ in terms of the partition function. Let us denote the probability of finding exactly $n$ particles in state $\alpha$ and $m$ particles in state $\gamma$ by $p_{\alpha\gamma}(n,m)$. This parameter may be calculated similarly to (\ref{oneprob}), using $p_{\alpha\gamma}^\geq(n,m)$ defined as the probability of finding at least $n$ particles in state $\alpha$ and at least $m$ in state $\gamma$. This quantity fulfills the following relation:
\begin{eqnarray}
p_{\alpha\gamma}(n,m)=p^\geq _{\alpha\gamma}(n,m)+p^\geq _{\alpha\gamma}(n+1,m+1)-\nonumber\\
-p^\geq _{\alpha\gamma}(n,m+1)-p^\geq _{\alpha\gamma}(n+1,m)
\label{pppp}
\end{eqnarray}
Similar calculations as for the mean occupation yields
\begin{equation}
p^\geq _{\alpha\gamma} (n,m)=e^{-n\beta\epsilon_\alpha-m\beta\epsilon_\gamma}\frac{Z(\beta,N-m-n)}{Z(\beta,N)}
\end{equation}
Next, we calculate $\left\langle n_\alpha n_\gamma \right\rangle$ from the definition: $\left\langle n_\alpha n_\gamma \right\rangle=\sum_{n,m}{n\,m\,p_{\alpha\gamma}(n,m)}$ and using \eqref{pppp}. After straightforward calculations, we get
\begin{equation}
\left\langle n_\alpha n_\gamma \right\rangle=\frac{1}{Z(\beta,N)}\sum_{n=1}^N{}\sum_{m=1}^N{e^{-n\beta\epsilon_\alpha-m\beta\epsilon_\gamma}Z(\beta,N-m-n)}
\label{correlations}
\end{equation}
Now, we are in position to calculate the mean occupation numbers and correlations between different sites of the optical lattices. Formally, correlations between two lattice sites are given by the trace of density matrix $\hat{\rho}$ with operators $\hat{n}_i \hat{n}_j$. For the canonical ensemble $\hat{\rho}=e^{-\beta \hat{H}}/Z(\beta,N)$, so $\langle n_i n_j\rangle = Z^{-1}(\beta,N)\textrm{Tr}\left\{e^{-\beta \hat{H}}\hat{n}_i \hat{n}_j\right\}$. Because $\hat{\rho}$ has a simple form only in the basis of hamiltonian eigenstates, we have to find the relation between creation and annihilation operators of single-particle eigenstates $\hat{b}_\alpha$ and lattice sites $\hat{g}_i$ numerically. By expressing the mean values of site operators by mean values of eigenstate operators and utilizing the fact that the number of atoms is constant so only certain terms are non-zero, we obtain
\begin{displaymath}
\left\langle n_i n_j \right\rangle = \sum_{\alpha}{\left|c^\alpha_i\right|^2\left|c^\alpha_j\right|^2 \left\langle n_\alpha^2\right\rangle}+ \sum_{\alpha\neq\eta}{\left|c^\alpha_i\right|^2\left|c^\eta_j\right|^2 \left\langle n_\alpha n_\eta\right\rangle}+
\end{displaymath}
\begin{displaymath}
+\sum_{\alpha\neq\eta}{c^{\alpha\star}_i c^\eta_i c^{\eta\star}_i c^\alpha_j \left\langle n_\alpha n_\eta\right\rangle}+\sum_{\alpha\neq\eta}{c^{\alpha\star}_i c^\eta_i c^{\eta\star}_i c^\alpha_j \left\langle n_\alpha\right\rangle}=
\end{displaymath}
\begin{equation}
=\sum_{\alpha,\eta}{\left|c^\alpha_i\right|^2\left|c^\eta_j\right|^2 \left\langle n_\alpha n_\eta\right\rangle}+\sum_{\alpha\neq\eta}{c^{\alpha\star}_i c^\eta_i c^{\eta\star}_i c^\alpha_j (\left\langle n_\alpha n_\eta\right\rangle+\left\langle n_\alpha \right\rangle)},
\label{sitefl}
\end{equation}
as well as the formula for the mean occupation numbers
\begin{equation}
\left\langle n_i \right\rangle = \sum_{\alpha}{\left|c^{\alpha}_i\right|^2 \left\langle n_\alpha\right\rangle}.
\end{equation}

\section{Ideal Bose gas with quadratic energy spectrum}
Let us consider a gas of noninteracting bosons in an external trap with the quadratic energy spectrum: $\epsilon_k = a k^2$, where $a$ is constant. The logarithm of the grand canonical partition function is by definition given by
\begin{equation}
\ln\Xi(z,\beta)=\sum_{k=0}^\infty{\ln\left(1-ze^{-\beta a k^2}\right)},
\end{equation}
where $z=e^{-\beta \mu}$ is called the fugacity. The expectation value of the total number of particles is given by $\langle N\rangle = z\frac{\partial}{\partial z} \ln\Xi(z,\beta)$, which yields
\begin{equation}
\langle N\rangle = \sum_{k=0}^\infty{\frac{ze^{-\beta a k^2}}{1-ze^{-\beta a k^2}}}=\sum_{k=0}^\infty{\frac{e^{\beta(\mu-ak^2)/2}}{2\sinh\left((\beta a k^2 - \beta \mu\right)/2)}}.
\end{equation}
In the low-temperature limit: $\beta \mu \ll 1$ and $\beta a \ll 1$, we may keep the first term in series expansion of $\sinh()$ and $\exp{}$ functions, obtaining
\begin{equation}
\langle N\rangle \approx \frac{z}{1-z}+\frac{1}{a\beta}\sum_{k=1}^\infty{\frac{1}{k^2}}=\frac{z}{1-z}+\frac{\pi^2 k_B T}{6a}.
\end{equation}
The ground state occupation $\langle N_0\rangle$ is equal to $z/(1-z)$. At the characteristic temperature $T_c$ it becomes macroscopic. From this observation we conclude that the characteristic temperature of the gas is given by $T_c=6aN/\pi^2$, and that below $T_c$
\begin{equation}
\frac{\langle N_0\rangle}{\langle N\rangle} \approx 1-\frac{T}{T_c}.
\end{equation}
\section{The partition function for the strongly interacting gas}
For the strongly interacting gas the previous method does not work but we may use the fact that the hamiltonian separates the lattice sites and we can develop a different method. We start from definition
\begin{equation}
Z(N,\beta)=\sum_{n_1}{}\ldots\sum_{n_M}{e^{-\beta\sum{\varepsilon_i n_i}-\beta U\sum{n_i(n_i-1)}/2}\delta_{\sum{n_i},N}}.
\end{equation}
Due to the separation of the lattice sites the partition function for the first $m$ lattice sites $Z(N,\beta,m)$ can be expressed by $Z(N,\beta,m-1)$:
\begin{equation}
Z_N(\beta,m)=\sum_{n_m}{e^{-\beta(\varepsilon_m n_m- \frac{U}{2}n_m(n_m-1))}Z_{N-n_m}(\beta,m-1)}.
\end{equation}
Calculation of the occupation numbers, fluctuations and correlations may be done in a similar manner.
\bibliography{Articles}

\begin{thebibliography}{43}
\expandafter\ifx\csname natexlab\endcsname\relax\def\natexlab#1{#1}\fi
\expandafter\ifx\csname bibnamefont\endcsname\relax
  \def\bibnamefont#1{#1}\fi
\expandafter\ifx\csname bibfnamefont\endcsname\relax
  \def\bibfnamefont#1{#1}\fi
\expandafter\ifx\csname citenamefont\endcsname\relax
  \def\citenamefont#1{#1}\fi
\expandafter\ifx\csname url\endcsname\relax
  \def\url#1{\texttt{#1}}\fi
\expandafter\ifx\csname urlprefix\endcsname\relax\def\urlprefix{URL }\fi
\providecommand{\bibinfo}[2]{#2}
\providecommand{\eprint}[2][]{\url{#2}}

\bibitem[{\citenamefont{Anderson}(1958)}]{Anderson1958}
\bibinfo{author}{\bibfnamefont{P.~W.} \bibnamefont{Anderson}},
  \bibinfo{journal}{Phys. Rev.} \textbf{\bibinfo{volume}{109}},
  \bibinfo{pages}{1492} (\bibinfo{year}{1958}).

\bibitem[{\citenamefont{Hefei et~al.}(2008)\citenamefont{Hefei, Strybulevych,
  Page, Skipetrov, and van Tiggelen}}]{Tiggelen}
\bibinfo{author}{\bibfnamefont{H.}~\bibnamefont{Hefei}},
  \bibinfo{author}{\bibfnamefont{A.}~\bibnamefont{Strybulevych}},
  \bibinfo{author}{\bibfnamefont{J.~H.} \bibnamefont{Page}},
  \bibinfo{author}{\bibfnamefont{S.~E.} \bibnamefont{Skipetrov}},
  \bibnamefont{and} \bibinfo{author}{\bibfnamefont{B.~A.} \bibnamefont{van
  Tiggelen}}, \bibinfo{journal}{Nat. Phys.} \textbf{\bibinfo{volume}{4}}
  (\bibinfo{year}{2008}).

\bibitem[{\citenamefont{Chabanov et~al.}(2000)\citenamefont{Chabanov,
  Stoytchev, and Genack}}]{Chabanov}
\bibinfo{author}{\bibfnamefont{A.~A.} \bibnamefont{Chabanov}},
  \bibinfo{author}{\bibfnamefont{M.}~\bibnamefont{Stoytchev}},
  \bibnamefont{and} \bibinfo{author}{\bibfnamefont{A.~Z.}
  \bibnamefont{Genack}}, \bibinfo{journal}{Nature}
  \textbf{\bibinfo{volume}{404}}, \bibinfo{pages}{850} (\bibinfo{year}{2000}).

\bibitem[{\citenamefont{Wiersma et~al.}(1997)\citenamefont{Wiersma, Bartolini,
  Lagendijk, and Righini}}]{Wiersma}
\bibinfo{author}{\bibfnamefont{D.~S.} \bibnamefont{Wiersma}},
  \bibinfo{author}{\bibfnamefont{P.}~\bibnamefont{Bartolini}},
  \bibinfo{author}{\bibfnamefont{A.}~\bibnamefont{Lagendijk}},
  \bibnamefont{and} \bibinfo{author}{\bibfnamefont{R.}~\bibnamefont{Righini}},
  \bibinfo{journal}{Nature} \textbf{\bibinfo{volume}{390}},
  \bibinfo{pages}{671} (\bibinfo{year}{1997}).

\bibitem[{\citenamefont{Roati et~al.}(2008)\citenamefont{Roati, D'Errico,
  Fallani, Fattori, Fort, Zaccanti, Modugno, Modugno, and Inguscio}}]{Inguscio}
\bibinfo{author}{\bibfnamefont{G.}~\bibnamefont{Roati}},
  \bibinfo{author}{\bibfnamefont{C.}~\bibnamefont{D'Errico}},
  \bibinfo{author}{\bibfnamefont{L.}~\bibnamefont{Fallani}},
  \bibinfo{author}{\bibfnamefont{M.}~\bibnamefont{Fattori}},
  \bibinfo{author}{\bibfnamefont{C.}~\bibnamefont{Fort}},
  \bibinfo{author}{\bibfnamefont{M.}~\bibnamefont{Zaccanti}},
  \bibinfo{author}{\bibfnamefont{G.}~\bibnamefont{Modugno}},
  \bibinfo{author}{\bibfnamefont{M.}~\bibnamefont{Modugno}}, \bibnamefont{and}
  \bibinfo{author}{\bibfnamefont{M.}~\bibnamefont{Inguscio}},
  \bibinfo{journal}{Nature} \textbf{\bibinfo{volume}{453}},
  \bibinfo{pages}{895} (\bibinfo{year}{2008}).

\bibitem[{\citenamefont{Billy et~al.}(2008)\citenamefont{Billy, Josse, Zuo,
  Bernard, Hambrecht, Lugan, Clement, Sanchez-Palencia, Bouyer, and
  Aspect}}]{Aspect}
\bibinfo{author}{\bibfnamefont{J.}~\bibnamefont{Billy}},
  \bibinfo{author}{\bibfnamefont{V.}~\bibnamefont{Josse}},
  \bibinfo{author}{\bibfnamefont{Z.}~\bibnamefont{Zuo}},
  \bibinfo{author}{\bibfnamefont{A.}~\bibnamefont{Bernard}},
  \bibinfo{author}{\bibfnamefont{B.}~\bibnamefont{Hambrecht}},
  \bibinfo{author}{\bibfnamefont{P.}~\bibnamefont{Lugan}},
  \bibinfo{author}{\bibfnamefont{D.}~\bibnamefont{Clement}},
  \bibinfo{author}{\bibfnamefont{L.}~\bibnamefont{Sanchez-Palencia}},
  \bibinfo{author}{\bibfnamefont{P.}~\bibnamefont{Bouyer}}, \bibnamefont{and}
  \bibinfo{author}{\bibfnamefont{A.}~\bibnamefont{Aspect}},
  \bibinfo{journal}{Nature} \textbf{\bibinfo{volume}{453}},
  \bibinfo{pages}{891} (\bibinfo{year}{2008}).

\bibitem[{\citenamefont{Bloch et~al.}(2008)\citenamefont{Bloch, Dalibard, and
  Zwerger}}]{Bloch2008}
\bibinfo{author}{\bibfnamefont{I.}~\bibnamefont{Bloch}},
  \bibinfo{author}{\bibfnamefont{J.}~\bibnamefont{Dalibard}}, \bibnamefont{and}
  \bibinfo{author}{\bibfnamefont{W.}~\bibnamefont{Zwerger}},
  \bibinfo{journal}{Rev. Mod. Phys.} \textbf{\bibinfo{volume}{80}},
  \bibinfo{eid}{885} (\bibinfo{year}{2008}).

\bibitem[{\citenamefont{Cl\'ement et~al.}(2005)\citenamefont{Cl\'ement,
  Var\'on, Hugbart, Retter, Bouyer, Sanchez-Palencia, Gangardt, Shlyapnikov,
  and Aspect}}]{AspectOld}
\bibinfo{author}{\bibfnamefont{D.}~\bibnamefont{Cl\'ement}},
  \bibinfo{author}{\bibfnamefont{A.~F.} \bibnamefont{Var\'on}},
  \bibinfo{author}{\bibfnamefont{M.}~\bibnamefont{Hugbart}},
  \bibinfo{author}{\bibfnamefont{J.~A.} \bibnamefont{Retter}},
  \bibinfo{author}{\bibfnamefont{P.}~\bibnamefont{Bouyer}},
  \bibinfo{author}{\bibfnamefont{L.}~\bibnamefont{Sanchez-Palencia}},
  \bibinfo{author}{\bibfnamefont{D.~M.} \bibnamefont{Gangardt}},
  \bibinfo{author}{\bibfnamefont{G.~V.} \bibnamefont{Shlyapnikov}},
  \bibnamefont{and} \bibinfo{author}{\bibfnamefont{A.}~\bibnamefont{Aspect}},
  \bibinfo{journal}{Phys. Rev. Lett.} \textbf{\bibinfo{volume}{95}},
  \bibinfo{pages}{170409} (\bibinfo{year}{2005}).

\bibitem[{\citenamefont{Clement et~al.}(2006)\citenamefont{Clement, Varen,
  Retter, Sanchez-Palencia, Aspect, and Bouyer}}]{AspectJourn}
\bibinfo{author}{\bibfnamefont{D.}~\bibnamefont{Clement}},
  \bibinfo{author}{\bibfnamefont{A.~F.} \bibnamefont{Varen}},
  \bibinfo{author}{\bibfnamefont{J.~A.} \bibnamefont{Retter}},
  \bibinfo{author}{\bibfnamefont{L.}~\bibnamefont{Sanchez-Palencia}},
  \bibinfo{author}{\bibfnamefont{A.}~\bibnamefont{Aspect}}, \bibnamefont{and}
  \bibinfo{author}{\bibfnamefont{P.}~\bibnamefont{Bouyer}},
  \bibinfo{journal}{New Journal of Physics} \textbf{\bibinfo{volume}{8}},
  \bibinfo{pages}{165} (\bibinfo{year}{2006}).

\bibitem[{\citenamefont{Lye et~al.}(2007)\citenamefont{Lye, Fallani, Fort,
  Guarrera, Modugno, Wiersma, and Inguscio}}]{InguscioPRA}
\bibinfo{author}{\bibfnamefont{J.~E.} \bibnamefont{Lye}},
  \bibinfo{author}{\bibfnamefont{L.}~\bibnamefont{Fallani}},
  \bibinfo{author}{\bibfnamefont{C.}~\bibnamefont{Fort}},
  \bibinfo{author}{\bibfnamefont{V.}~\bibnamefont{Guarrera}},
  \bibinfo{author}{\bibfnamefont{M.}~\bibnamefont{Modugno}},
  \bibinfo{author}{\bibfnamefont{D.~S.} \bibnamefont{Wiersma}},
  \bibnamefont{and} \bibinfo{author}{\bibfnamefont{M.}~\bibnamefont{Inguscio}},
  \bibinfo{journal}{Phys. Rev. A} \textbf{\bibinfo{volume}{75}},
  \bibinfo{pages}{061603} (\bibinfo{year}{2007}).

\bibitem[{\citenamefont{Guarrera et~al.}(2008)\citenamefont{Guarrera, Fabbri,
  Fallani, Fort, van~der Stam, and Inguscio}}]{Guarrera2007}
\bibinfo{author}{\bibfnamefont{V.}~\bibnamefont{Guarrera}},
  \bibinfo{author}{\bibfnamefont{N.}~\bibnamefont{Fabbri}},
  \bibinfo{author}{\bibfnamefont{L.}~\bibnamefont{Fallani}},
  \bibinfo{author}{\bibfnamefont{C.}~\bibnamefont{Fort}},
  \bibinfo{author}{\bibfnamefont{K.~M.~R.} \bibnamefont{van~der Stam}},
  \bibnamefont{and} \bibinfo{author}{\bibfnamefont{M.}~\bibnamefont{Inguscio}},
  \bibinfo{journal}{Phys. Rev. Lett.} \textbf{\bibinfo{volume}{100}},
  \bibinfo{pages}{250403} (\bibinfo{year}{2008}).

\bibitem[{\citenamefont{Thouless}(1974)}]{Thouless}
\bibinfo{author}{\bibfnamefont{D.}~\bibnamefont{Thouless}},
  \bibinfo{journal}{Physics Reports} \textbf{\bibinfo{volume}{13}},
  \bibinfo{pages}{93} (\bibinfo{year}{1974}).

\bibitem[{\citenamefont{Anderson et~al.}(1980)\citenamefont{Anderson, Thouless,
  Abrahams, and Fisher}}]{Gang4}
\bibinfo{author}{\bibfnamefont{P.~W.} \bibnamefont{Anderson}},
  \bibinfo{author}{\bibfnamefont{D.~J.} \bibnamefont{Thouless}},
  \bibinfo{author}{\bibfnamefont{E.}~\bibnamefont{Abrahams}}, \bibnamefont{and}
  \bibinfo{author}{\bibfnamefont{D.~S.} \bibnamefont{Fisher}},
  \bibinfo{journal}{Phys. Rev. B} \textbf{\bibinfo{volume}{22}},
  \bibinfo{pages}{3519} (\bibinfo{year}{1980}).

\bibitem[{\citenamefont{von Dreifus and Klein}(1989)}]{Proof}
\bibinfo{author}{\bibfnamefont{H.}~\bibnamefont{von Dreifus}} \bibnamefont{and}
  \bibinfo{author}{\bibfnamefont{A.}~\bibnamefont{Klein}},
  \bibinfo{journal}{Communications in Mathematical Physics}
  \textbf{\bibinfo{volume}{124}}, \bibinfo{pages}{285} (\bibinfo{year}{1989}),
  \bibinfo{note}{10.1007/BF01219198}.

\bibitem[{\citenamefont{Aubry and Andre}(1980)}]{Aubry1980}
\bibinfo{author}{\bibfnamefont{S.}~\bibnamefont{Aubry}} \bibnamefont{and}
  \bibinfo{author}{\bibfnamefont{G.}~\bibnamefont{Andre}},
  \bibinfo{journal}{Ann. Isr. Phys. Soc.} \textbf{\bibinfo{volume}{3}},
  \bibinfo{pages}{33} (\bibinfo{year}{1980}).

\bibitem[{\citenamefont{Fisher et~al.}(1989)\citenamefont{Fisher, Weichman,
  Grinstein, and Fisher}}]{Fisher1989}
\bibinfo{author}{\bibfnamefont{M.~P.~A.} \bibnamefont{Fisher}},
  \bibinfo{author}{\bibfnamefont{P.~B.} \bibnamefont{Weichman}},
  \bibinfo{author}{\bibfnamefont{G.}~\bibnamefont{Grinstein}},
  \bibnamefont{and} \bibinfo{author}{\bibfnamefont{D.~S.}
  \bibnamefont{Fisher}}, \bibinfo{journal}{Phys. Rev. B}
  \textbf{\bibinfo{volume}{40}}, \bibinfo{pages}{546} (\bibinfo{year}{1989}).

\bibitem[{\citenamefont{Damski et~al.}(2003)\citenamefont{Damski, Zakrzewski,
  Santos, Zoller, and Lewenstein}}]{Lewenstein2003}
\bibinfo{author}{\bibfnamefont{B.}~\bibnamefont{Damski}},
  \bibinfo{author}{\bibfnamefont{J.}~\bibnamefont{Zakrzewski}},
  \bibinfo{author}{\bibfnamefont{L.}~\bibnamefont{Santos}},
  \bibinfo{author}{\bibfnamefont{P.}~\bibnamefont{Zoller}}, \bibnamefont{and}
  \bibinfo{author}{\bibfnamefont{M.}~\bibnamefont{Lewenstein}},
  \bibinfo{journal}{Phys. Rev. Lett.} \textbf{\bibinfo{volume}{91}},
  \bibinfo{pages}{080403} (\bibinfo{year}{2003}).

\bibitem[{\citenamefont{Roux et~al.}(2008)\citenamefont{Roux, Barthel,
  McCulloch, Kollath, Schollw\"ock, and Giamarchi}}]{Roux2008}
\bibinfo{author}{\bibfnamefont{G.}~\bibnamefont{Roux}},
  \bibinfo{author}{\bibfnamefont{T.}~\bibnamefont{Barthel}},
  \bibinfo{author}{\bibfnamefont{I.~P.} \bibnamefont{McCulloch}},
  \bibinfo{author}{\bibfnamefont{C.}~\bibnamefont{Kollath}},
  \bibinfo{author}{\bibfnamefont{U.}~\bibnamefont{Schollw\"ock}},
  \bibnamefont{and}
  \bibinfo{author}{\bibfnamefont{T.}~\bibnamefont{Giamarchi}},
  \bibinfo{journal}{Phys. Rev. A} \textbf{\bibinfo{volume}{78}},
  \bibinfo{pages}{023628} (\bibinfo{year}{2008}).

\bibitem[{\citenamefont{Fontanesi et~al.}(2010)\citenamefont{Fontanesi,
  Wouters, and Savona}}]{Fontanesi2008}
\bibinfo{author}{\bibfnamefont{L.}~\bibnamefont{Fontanesi}},
  \bibinfo{author}{\bibfnamefont{M.}~\bibnamefont{Wouters}}, \bibnamefont{and}
  \bibinfo{author}{\bibfnamefont{V.}~\bibnamefont{Savona}},
  \bibinfo{journal}{Phys. Rev. A} \textbf{\bibinfo{volume}{81}},
  \bibinfo{pages}{053603} (\bibinfo{year}{2010}).

\bibitem[{\citenamefont{Greiner et~al.}(2002)\citenamefont{Greiner, Mandel,
  Esslinger, Hansch, and Bloch}}]{BlochNature}
\bibinfo{author}{\bibfnamefont{M.}~\bibnamefont{Greiner}},
  \bibinfo{author}{\bibfnamefont{O.}~\bibnamefont{Mandel}},
  \bibinfo{author}{\bibfnamefont{T.}~\bibnamefont{Esslinger}},
  \bibinfo{author}{\bibfnamefont{T.}~\bibnamefont{Hansch}}, \bibnamefont{and}
  \bibinfo{author}{\bibfnamefont{I.}~\bibnamefont{Bloch}},
  \bibinfo{journal}{Nature} \textbf{\bibinfo{volume}{415}}, \bibinfo{pages}{39}
  (\bibinfo{year}{2002}).

\bibitem[{\citenamefont{Roth and Burnett}(2003)}]{Burnett2003}
\bibinfo{author}{\bibfnamefont{R.}~\bibnamefont{Roth}} \bibnamefont{and}
  \bibinfo{author}{\bibfnamefont{K.}~\bibnamefont{Burnett}},
  \bibinfo{journal}{Phys. Rev. A} \textbf{\bibinfo{volume}{68}},
  \bibinfo{pages}{023604} (\bibinfo{year}{2003}).

\bibitem[{\citenamefont{Folling et~al.}(2005)\citenamefont{Folling, Gerbier,
  Widera, Mandel, Gericke, and Bloch}}]{BlochNoise}
\bibinfo{author}{\bibfnamefont{S.}~\bibnamefont{Folling}},
  \bibinfo{author}{\bibfnamefont{F.}~\bibnamefont{Gerbier}},
  \bibinfo{author}{\bibfnamefont{A.}~\bibnamefont{Widera}},
  \bibinfo{author}{\bibfnamefont{O.}~\bibnamefont{Mandel}},
  \bibinfo{author}{\bibfnamefont{T.}~\bibnamefont{Gericke}}, \bibnamefont{and}
  \bibinfo{author}{\bibfnamefont{I.}~\bibnamefont{Bloch}},
  \bibinfo{journal}{Nature} \textbf{\bibinfo{volume}{453}},
  \bibinfo{pages}{481} (\bibinfo{year}{2005}).

\bibitem[{\citenamefont{Altman et~al.}(2004)\citenamefont{Altman, Demler, and
  Lukin}}]{Demler}
\bibinfo{author}{\bibfnamefont{E.}~\bibnamefont{Altman}},
  \bibinfo{author}{\bibfnamefont{E.}~\bibnamefont{Demler}}, \bibnamefont{and}
  \bibinfo{author}{\bibfnamefont{M.~D.} \bibnamefont{Lukin}},
  \bibinfo{journal}{Phys. Rev. A} \textbf{\bibinfo{volume}{70}},
  \bibinfo{pages}{013603} (\bibinfo{year}{2004}).

\bibitem[{\citenamefont{Weitenberg et~al.}(2011)\citenamefont{Weitenberg,
  Schau\ss{}, Fukuhara, Cheneau, Endres, Bloch, and Kuhr}}]{Bloch2011}
\bibinfo{author}{\bibfnamefont{C.}~\bibnamefont{Weitenberg}},
  \bibinfo{author}{\bibfnamefont{P.}~\bibnamefont{Schau\ss{}}},
  \bibinfo{author}{\bibfnamefont{T.}~\bibnamefont{Fukuhara}},
  \bibinfo{author}{\bibfnamefont{M.}~\bibnamefont{Cheneau}},
  \bibinfo{author}{\bibfnamefont{M.}~\bibnamefont{Endres}},
  \bibinfo{author}{\bibfnamefont{I.}~\bibnamefont{Bloch}}, \bibnamefont{and}
  \bibinfo{author}{\bibfnamefont{S.}~\bibnamefont{Kuhr}},
  \bibinfo{journal}{Phys. Rev. Lett.} \textbf{\bibinfo{volume}{106}},
  \bibinfo{pages}{215301} (\bibinfo{year}{2011}).

\bibitem[{\citenamefont{Lewenstein and You}(1993)}]{Lewenstein1993}
\bibinfo{author}{\bibfnamefont{M.}~\bibnamefont{Lewenstein}} \bibnamefont{and}
  \bibinfo{author}{\bibfnamefont{L.}~\bibnamefont{You}},
  \bibinfo{journal}{Phys. Rev. Lett.} \textbf{\bibinfo{volume}{71}},
  \bibinfo{pages}{1339} (\bibinfo{year}{1993}).

\bibitem[{\citenamefont{Idziaszek et~al.}(2000)\citenamefont{Idziaszek,
  Rza\ifmmode \mbox{\c{}}\else \c{}\fi{}\ifmmode~\dot{z}\else \.{z}\fi{}ewski,
  and Lewenstein}}]{Idziaszek}
\bibinfo{author}{\bibfnamefont{Z.}~\bibnamefont{Idziaszek}},
  \bibinfo{author}{\bibfnamefont{K.}~\bibnamefont{Rza\ifmmode \mbox{\c{}}\else
  \c{}\fi{}\ifmmode~\dot{z}\else \.{z}\fi{}ewski}}, \bibnamefont{and}
  \bibinfo{author}{\bibfnamefont{M.}~\bibnamefont{Lewenstein}},
  \bibinfo{journal}{Phys. Rev. A} \textbf{\bibinfo{volume}{61}},
  \bibinfo{pages}{053608} (\bibinfo{year}{2000}).

\bibitem[{\citenamefont{Zhang et~al.}(1999)\citenamefont{Zhang, Sackett, and
  Hulet}}]{Zhang1999}
\bibinfo{author}{\bibfnamefont{W.}~\bibnamefont{Zhang}},
  \bibinfo{author}{\bibfnamefont{C.~A.} \bibnamefont{Sackett}},
  \bibnamefont{and} \bibinfo{author}{\bibfnamefont{R.~G.} \bibnamefont{Hulet}},
  \bibinfo{journal}{Phys. Rev. A} \textbf{\bibinfo{volume}{60}},
  \bibinfo{pages}{504} (\bibinfo{year}{1999}).

\bibitem[{\citenamefont{Mekhov et~al.}(2007)\citenamefont{Mekhov, Maschler, and
  Ritsch}}]{Mekhov2007}
\bibinfo{author}{\bibfnamefont{I.}~\bibnamefont{Mekhov}},
  \bibinfo{author}{\bibfnamefont{C.}~\bibnamefont{Maschler}}, \bibnamefont{and}
  \bibinfo{author}{\bibfnamefont{H.}~\bibnamefont{Ritsch}},
  \bibinfo{journal}{Nat. Phys.} \textbf{\bibinfo{volume}{3}},
  \bibinfo{pages}{319} (\bibinfo{year}{2007}).

\bibitem[{\citenamefont{\L{}akomy et~al.}(2009)\citenamefont{\L{}akomy,
  Idziaszek, and Trippenbach}}]{Lakomy}
\bibinfo{author}{\bibfnamefont{K.}~\bibnamefont{\L{}akomy}},
  \bibinfo{author}{\bibfnamefont{Z.}~\bibnamefont{Idziaszek}},
  \bibnamefont{and}
  \bibinfo{author}{\bibfnamefont{M.}~\bibnamefont{Trippenbach}},
  \bibinfo{journal}{Phys. Rev. A} \textbf{\bibinfo{volume}{80}},
  \bibinfo{pages}{043404} (\bibinfo{year}{2009}).

\bibitem[{\citenamefont{Rist et~al.}(2010)\citenamefont{Rist, Menotti, and
  Morigi}}]{Menotti2010}
\bibinfo{author}{\bibfnamefont{S.}~\bibnamefont{Rist}},
  \bibinfo{author}{\bibfnamefont{C.}~\bibnamefont{Menotti}}, \bibnamefont{and}
  \bibinfo{author}{\bibfnamefont{G.}~\bibnamefont{Morigi}},
  \bibinfo{journal}{Phys. Rev. A} \textbf{\bibinfo{volume}{81}},
  \bibinfo{pages}{013404} (\bibinfo{year}{2010}).

\bibitem[{\citenamefont{Douglas and Burnett}(2011)}]{Douglas2011}
\bibinfo{author}{\bibfnamefont{J.~S.} \bibnamefont{Douglas}} \bibnamefont{and}
  \bibinfo{author}{\bibfnamefont{K.}~\bibnamefont{Burnett}},
  \bibinfo{journal}{Phys. Rev. A} \textbf{\bibinfo{volume}{84}},
  \bibinfo{pages}{033637} (\bibinfo{year}{2011}).

\bibitem[{\citenamefont{Stenger et~al.}(1999)\citenamefont{Stenger, Inouye,
  Chikkatur, Stamper-Kurn, Pritchard, and Ketterle}}]{Bragg}
\bibinfo{author}{\bibfnamefont{J.}~\bibnamefont{Stenger}},
  \bibinfo{author}{\bibfnamefont{S.}~\bibnamefont{Inouye}},
  \bibinfo{author}{\bibfnamefont{A.~P.} \bibnamefont{Chikkatur}},
  \bibinfo{author}{\bibfnamefont{D.~M.} \bibnamefont{Stamper-Kurn}},
  \bibinfo{author}{\bibfnamefont{D.~E.} \bibnamefont{Pritchard}},
  \bibnamefont{and} \bibinfo{author}{\bibfnamefont{W.}~\bibnamefont{Ketterle}},
  \bibinfo{journal}{Phys. Rev. Lett.} \textbf{\bibinfo{volume}{82}},
  \bibinfo{pages}{4569} (\bibinfo{year}{1999}).

\bibitem[{\citenamefont{Eckert et~al.}(2008)\citenamefont{Eckert, Romero-Isart,
  Rodriguez, Lewenstein, Polzik, and Sanpera}}]{LewensteinNature}
\bibinfo{author}{\bibfnamefont{K.}~\bibnamefont{Eckert}},
  \bibinfo{author}{\bibfnamefont{O.}~\bibnamefont{Romero-Isart}},
  \bibinfo{author}{\bibfnamefont{M.}~\bibnamefont{Rodriguez}},
  \bibinfo{author}{\bibfnamefont{M.}~\bibnamefont{Lewenstein}},
  \bibinfo{author}{\bibfnamefont{E.}~\bibnamefont{Polzik}}, \bibnamefont{and}
  \bibinfo{author}{\bibfnamefont{A.}~\bibnamefont{Sanpera}},
  \bibinfo{journal}{Nat. Phys.} \textbf{\bibinfo{volume}{4}},
  \bibinfo{pages}{50} (\bibinfo{year}{2008}).

\bibitem[{\citenamefont{Jaksch et~al.}(1998)\citenamefont{Jaksch, Bruder,
  Cirac, Gardiner, and Zoller}}]{Jaksch1998}
\bibinfo{author}{\bibfnamefont{D.}~\bibnamefont{Jaksch}},
  \bibinfo{author}{\bibfnamefont{C.}~\bibnamefont{Bruder}},
  \bibinfo{author}{\bibfnamefont{J.~I.} \bibnamefont{Cirac}},
  \bibinfo{author}{\bibfnamefont{C.~W.} \bibnamefont{Gardiner}},
  \bibnamefont{and} \bibinfo{author}{\bibfnamefont{P.}~\bibnamefont{Zoller}},
  \bibinfo{journal}{Phys. Rev. Lett.} \textbf{\bibinfo{volume}{81}},
  \bibinfo{pages}{3108} (\bibinfo{year}{1998}).

\bibitem[{\citenamefont{Modugno}(2009)}]{Modugno2009}
\bibinfo{author}{\bibfnamefont{M.}~\bibnamefont{Modugno}},
  \bibinfo{journal}{New Journal of Physics} \textbf{\bibinfo{volume}{11}},
  \bibinfo{pages}{033023} (\bibinfo{year}{2009}).

\bibitem[{\citenamefont{Aulbach et~al.}(2004)\citenamefont{Aulbach, Wobst,
  Ingold, Hänggi, and Varga}}]{Aulbach}
\bibinfo{author}{\bibfnamefont{C.}~\bibnamefont{Aulbach}},
  \bibinfo{author}{\bibfnamefont{A.}~\bibnamefont{Wobst}},
  \bibinfo{author}{\bibfnamefont{G.-L.} \bibnamefont{Ingold}},
  \bibinfo{author}{\bibfnamefont{P.}~\bibnamefont{Hänggi}}, \bibnamefont{and}
  \bibinfo{author}{\bibfnamefont{I.}~\bibnamefont{Varga}},
  \bibinfo{journal}{New Journal of Physics} \textbf{\bibinfo{volume}{6}},
  \bibinfo{pages}{70} (\bibinfo{year}{2004}).

\bibitem[{\citenamefont{Jitomirskaya}(1999)}]{Jitomirskaya1999}
\bibinfo{author}{\bibfnamefont{S.~Y.} \bibnamefont{Jitomirskaya}},
  \bibinfo{journal}{The Annals of Mathematics} \textbf{\bibinfo{volume}{150}},
  \bibinfo{pages}{pp. 1159} (\bibinfo{year}{1999}).

\bibitem[{\citenamefont{Politzer}(1996)}]{Politzer1996}
\bibinfo{author}{\bibfnamefont{H.~D.} \bibnamefont{Politzer}},
  \bibinfo{journal}{Phys. Rev. A} \textbf{\bibinfo{volume}{54}},
  \bibinfo{pages}{5048} (\bibinfo{year}{1996}).

\bibitem[{\citenamefont{Gajda and Rza\ifmmode \mbox{\c{}}\else
  \c{}\fi{}\ifmmode~\dot{z}\else \.{z}\fi{}ewski}(1997)}]{Gajda1997}
\bibinfo{author}{\bibfnamefont{M.}~\bibnamefont{Gajda}} \bibnamefont{and}
  \bibinfo{author}{\bibfnamefont{K.}~\bibnamefont{Rza\ifmmode \mbox{\c{}}\else
  \c{}\fi{}\ifmmode~\dot{z}\else \.{z}\fi{}ewski}}, \bibinfo{journal}{Phys.
  Rev. Lett.} \textbf{\bibinfo{volume}{78}}, \bibinfo{pages}{2686}
  (\bibinfo{year}{1997}).

\bibitem[{\citenamefont{Navez et~al.}(1997)\citenamefont{Navez, Bitouk, Gajda,
  Idziaszek, and Rza\ifmmode \mbox{\c{}}\else \c{}\fi{}\ifmmode~\dot{z}\else
  \.{z}\fi{}ewski}}]{Navez}
\bibinfo{author}{\bibfnamefont{P.}~\bibnamefont{Navez}},
  \bibinfo{author}{\bibfnamefont{D.}~\bibnamefont{Bitouk}},
  \bibinfo{author}{\bibfnamefont{M.}~\bibnamefont{Gajda}},
  \bibinfo{author}{\bibfnamefont{Z.}~\bibnamefont{Idziaszek}},
  \bibnamefont{and} \bibinfo{author}{\bibfnamefont{K.}~\bibnamefont{Rza\ifmmode
  \mbox{\c{}}\else \c{}\fi{}\ifmmode~\dot{z}\else \.{z}\fi{}ewski}},
  \bibinfo{journal}{Phys. Rev. Lett.} \textbf{\bibinfo{volume}{79}},
  \bibinfo{pages}{1789} (\bibinfo{year}{1997}).

\bibitem[{\citenamefont{Weiss and Wilkens}(1997)}]{Weiss97}
\bibinfo{author}{\bibfnamefont{C.}~\bibnamefont{Weiss}} \bibnamefont{and}
  \bibinfo{author}{\bibfnamefont{M.}~\bibnamefont{Wilkens}},
  \bibinfo{journal}{Opt. Express} \textbf{\bibinfo{volume}{1}},
  \bibinfo{pages}{272} (\bibinfo{year}{1997}).

\bibitem[{\citenamefont{Grossmann and Holthaus}(1997)}]{Grossmann97}
\bibinfo{author}{\bibfnamefont{S.}~\bibnamefont{Grossmann}} \bibnamefont{and}
  \bibinfo{author}{\bibfnamefont{M.}~\bibnamefont{Holthaus}},
  \bibinfo{journal}{Opt. Express} \textbf{\bibinfo{volume}{1}},
  \bibinfo{pages}{262} (\bibinfo{year}{1997}).

\bibitem[{\citenamefont{Idziaszek}(2002)}]{phd}
\bibinfo{author}{\bibfnamefont{Z.}~\bibnamefont{Idziaszek}}, Ph.D. thesis,
  \bibinfo{school}{Center for Theoretical Physics, Polish Academy of Sciences}
  (\bibinfo{year}{2002}).

\end{thebibliography}
\end{document}